\documentclass{aa}  
\usepackage{graphicx}
\usepackage{txfonts}

\usepackage{xcolor}
\usepackage{url}
\usepackage{soul} % YL : pour barrer du texte

\usepackage{nicefrac}

\usepackage{natbib}

\bibpunct{(}{)}{;}{a}{}{,} % to follow the A&A style
\begin{document} 
	
	\title{SPInS, a pipeline for massive stellar parameter inference}
	
	\subtitle{A public Python tool to age-date, weigh, size up stars, and more}
	
	\author{Y. Lebreton\inst{1,2} \and
		D. R. Reese\inst{1}
		%\fnmsep\thanks{Just to show the usage 			of the elements in the author field}
	}
	
	\institute{LESIA, Observatoire de Paris, Université PSL, CNRS, Sorbonne Université, Université de Paris, 5 place Jules Janssen, 92195 Meudon,
		\textsc{France}\\
		\email{yveline.lebreton@obspm.fr, daniel.reese@obspm.fr }
		\and
		Univ Rennes, CNRS, IPR (Institut de Physique de Rennes) - UMR 6251, F-35000 Rennes, \textsc{France}\\
		%\email{@obspm.fr}
		%\thanks{}
	}
	
	\date{Received 8 June 2020 ; accepted July 2020 } %(j'ai un petit doute sur la date ... \YLD{très improbable en effet :)} \DRR{peut-être l'éditeur remarquera ;)})}
	
	% \abstract{}{}{}{}{} 
	% 5 {} token are mandatory
	
	\abstract
	{Stellar parameters are required in a variety of contexts, ranging from the characterisation of exoplanets to Galactic archaeology. Among them, the age of stars cannot be directly measured, while the mass and radius can be measured in some particular cases (e.g. binary systems, interferometry). More generally, stellar ages, masses, and radii have to be inferred from stellar evolution models by appropriate techniques.}
	{We have designed a Python tool named SPInS. It takes a set of photometric, spectroscopic, interferometric, and/or asteroseismic observational constraints and, relying on a stellar model grid, provides the age, mass, and radius of a star, among others, as well as error bars and correlations. We make the tool available to the community via a dedicated website.}
	{SPInS uses a Bayesian approach to find the probability distribution function of stellar parameters from a set of classical constraints. At the heart of the code is a 
		Markov Chain Monte Carlo 
		%MCMC 
		solver coupled with interpolation within a pre-computed {stellar model} grid. Priors can be considered, such as the initial mass function or stellar formation rate. SPInS can characterise  single stars or coeval stars, such as members of binary systems or of stellar clusters.}
	{We first illustrate the capabilities of SPInS by studying stars that are spread over the Hertzsprung-Russell diagram. We then validate the tool by inferring the ages and masses of stars in several catalogues and by  comparing them with literature results. We show that in addition to the age and mass, SPInS can efficiently provide derived quantities, such as the radius, surface gravity, and seismic indices. We demonstrate that SPInS can age-date and characterise coeval stars that share a common age and chemical composition.}
	{The SPInS tool will be very helpful in preparing and interpreting the results of large-scale surveys, such as the wealth of data expected or already provided by space missions, such as Gaia, Kepler, TESS, and PLATO.}
	
	\keywords{Stars: fundamental parameters -- Methods: numerical -- Hertzsprung-Russell and C-M diagrams -- Asteroseismology}
	
	\maketitle
	
	%-------------------------------------------------------------------
	
	\section{Introduction}

	Stellar ages, masses, and radii (hereafter stellar parameters) are indispensable basic inputs in many  astrophysical studies, such as the study of the chemo-kinematical structure of the Milky Way (i.e. Galactic archaeology), exoplanetology, and cosmology. Indeed, stellar parameters have long been used to answer questions on how stars populating the different structures, that is the discs, bulge, and halo, in our Galaxy were formed and evolve, and to decipher in-situ formation, migration, and mergers. In this context, stellar parameters are the basis of stellar age-metallicity and age-velocity relations, the stellar initial mass function (IMF), or the stellar formation rate (SFR) [see \citet[][]{2014EAS....65..349H} for a review]. 
	Also, the ages of the oldest stars provide a robust lower limit to the age of the Universe.   
	Recently, with the discovery of several thousands of exoplanetary systems, it has become evident that no characterisation of the internal structure and evolutionary stage of planets is possible without a precise determination of the radius, mass, and age of the host stars \citep[see e.g.][]{2014ExA....38..249R}. 
	
	Today, the availability of observations from  large-scale astrometric, photometric,  spectroscopic, and interferometric surveys has made the demand for very precise and accurate stellar parameters acute.
	With Gaia \citep{2018A&A...616A...1G} and large spectroscopic surveys being conducted in parallel 
	%like the Gaia-ESO survey, RAVE, APOGEE, WEAVE, 4MOST, MOONS, GALAH
	\citep[see details and references in Sect. 2 of ][]{2017AN....338..644M}, the number of stars with precise astrometry, kinematics, and abundances will increase by more than three orders of magnitude. With high-precision photometry space-borne missions such as CoRoT \citep{2006cosp...36.3749B}, Kepler \citep{2010Sci...327..977B}, K2 \citep{2014PASP..126..398H}, TESS \citep{2015JATIS...1a4003R}, and in the future PLATO \citep{2014ExA....38..249R}, thousands of exoplanets have been and will be discovered. For planetary host stars of F, G, K spectral type on the main sequence or on the red giant branch, it is and will be possible to extract the power spectrum of their solar-like oscillations from the observed light curve, providing asteroseismic constraints to their modelling. Asteroseismology, therefore, will give access to precise and accurate masses, radii, and ages for these stars \citep[see e.g.][]{2014A&A...569A..21L, 2015MNRAS.452.2127S}. However, for cold M-type stars, which are optimal candidates for hosting habitable planets, the availability of asteroseismic constraints is less probable. 
	In this context, to fully exploit these rich data harvests and ensure scientific returns, we need modern numerical tools that are able to infer the stellar parameters of very large samples of stars. 	
	
	\citet{2010ARA&A..48..581S} reviewed the various methods that can be applied to age-date a star,  pointing out that a given method  (i) in most cases is only applicable to a limited range of stellar masses or evolutionary stages, (ii) can provide either absolute or relative ages, and (iii) is sometimes only applicable to very small stellar samples. Moreover, the precision and accuracy tightly depend on the age-dating method.
	Here, we  focus on the so-called isochrone placement method \citep{1993A&A...275..101E} that has long been used for age-dating and weighing stars in extended regions of
	the Hertzsprung-Russell  (hereafter H--R) diagram. 
	This method only requires having stellar evolutionary models available and is rather straightforward. It can provide ages and masses when other more powerful techniques, such as asteroseismology, are not applicable. It can also serve as a  reference when several age-dating methods are applicable. The precision depends on the star's mass and evolutionary state. Basically, the method consists in inferring the age and mass of an observed star with measured effective temperature, absolute magnitude, and metallicity (hereafter classical data) or any proxy for them, by looking for the theoretical stellar model that best fits the observations 
	\citep[e.g. ][]{1993A&A...275..101E,1998A&A...329..943N}.
	The adjustment can be performed in different ways. The simplest way to proceed is to select the appropriate isochrone by a $\chi^2$-minimisation, that is by searching among the isochrone points which one is closest to the star's location, in the related parameter space \citep[see e.g.][]{1998A&A...329..943N}. However, the selection of the right isochrone may be difficult in regions of the H--R diagram where they have a complex shape. In such regions, the evolutionary state of the star cannot be determined unambiguously and the star's position can equally be fitted with several isochrone points of different ages, masses, and metal contents. To improve the age-dating procedure, \citet{2004MNRAS.351..487P} proposed a Bayesian approach that, by adding 	prior information about the stellar and Galactic properties, allows the procedure to choose the most probable age. 
	The technique has been refined and improved by \citet{2005A&A...436..127J}, \citet{2006A&A...458..609D}, \citet{2006ApJ...645.1436V}, \citet{2007ApJS..168..297T}, \citet{2008MNRAS.383.1603H}, and \citet{2011A&A...530A.138C}, for the earlier papers, and reviewed by \citet{2014EAS....65..225V} and \citet{2014EAS....65..267V}. 
	In this context, the work by \citet{2006A&A...458..609D} gave birth to the PARAM\footnote{PARAM:~\url{http://stev.oapd.inaf.it/cgi-bin/param}}  web interface for the Bayesian estimation of stellar parameters. Then, a few stellar age-dating public codes  with different specificities were made public:  BASE-9\footnote{BASE-9:~ \url{https://github.com/BayesianStellarEvolution
		}} that allows the users to infer the properties of stellar clusters and their members, including white dwarfs \citep{2006ApJ...645.1436V}, UniDAM\footnote{UniDAM:~  \url{http://www2.mps.mpg.de/homes/mints/unidam.html}}  that can be used to exploit large stellar surveys \citep{2019A&A...629A.127M},  {stardate}\footnote{stardate:~\url{https://github.com/RuthAngus/stardate}} that considers constraints from gyrochronology \citep{2019AJ....158..173A}, and MCMCI\footnote{MCMCI: ~ \url{https://github.com/dfm/exoplanet}} that is dedicated to the characterisation of exoplanetary systems \citep{2020A&A...635A...6B}.
	
	In this work, we present and make public\footnote{SPInS: \url{https://gitlab.obspm.fr/dreese/spins}} a new tool based on Python and Fortran, named SPInS (standing for Stellar Parameters Inferred Systematically). SPInS is a modified version of the AIMS (Asteroseismic Inference on a Massive Scale) pipeline\footnote{AIMS: \url{https://lesia.obspm.fr/perso/daniel-reese/spaceinn/aims/}}. The AIMS code has been described and evaluated by  \citet{2018ASSP...49..149L} and \citet{2019MNRAS.484..771R}. AIMS is able to perform a full asteroseismic analysis and can estimate stellar parameters from two sets of observations: 
	classical data and detailed asteroseismic constraints (individual oscillation frequencies or a combination thereof).  While AIMS is essentially an asteroseismic tool, SPInS is not intended to handle detailed seismic data but rather focuses on classical or mean observed stellar data (these will be explained in the following sections). This greatly simplifies the procedure with a substantial gain in computational time and occupied disc space. In particular, SPInS only uses the standard outputs of stellar evolution models and does not need to be provided with the detailed calculations of the oscillation spectrum of the models.
	
	SPInS was initially created in 2018 to be used in hands-on sessions during the 5$^{\mathrm{th}}$ International Young Astronomer School held in Paris\footnote{International School organised by the Paris Doctoral School of Astronomy \& Astrophysics, see  \url{https://gaiaschool.wixsite.com/gaia-school2018}.}. The goal of SPInS is to estimate stellar ages and masses, as well as other properties and their error bars, in a probabilistic manner.
	This tool takes in a grid of stellar evolutionary tracks and applies a Monte Carlo Markov Chain (MCMC) approach in combination with a multidimensional interpolation scheme in order to find which stellar model(s) best reproduce(s) the observed luminosity $L_\star$ (or any proxy for it, such as the absolute  magnitude in a given band $M_\mathrm{b, \star}$), effective temperature $T_\mathrm{eff, \star}$ (or any colour index), and observed surface metal content $[\mathrm{M/H}]$. The latter can be replaced or complemented by other data derived from observations, such as the surface gravity $\log g$, the mass or radius, or both (for stars in eclipsing, spectroscopic, visual binaries, or with interferometric measurements), or asteroseismic parameters (the frequency at maximum power, the mean large frequency separation inferred  from the pressure-mode power spectrum, etc.). The advantage of this approach is that it provides a full probability distribution function (hereafter PDF) for any stellar parameter to be inferred, thereby accounting for multiple solutions when present.  It also allows the user to incorporate in the calculation various priors (i.e. a priori assumptions), such as the initial mass function (IMF), the stellar formation rate (SFR), or the metallicity distribution function (MDF). SPInS can be used in two operational modes: characterisation of a single star or {characterisation} of coeval groups, including binaries and stellar clusters. SPInS  is mostly written in Python with a modular structure to facilitate contributions from the community. Only a few computationally intensive parts have been written in Fortran in order to speed up calculations.
	
	The paper is organised as follows. In Sect. \ref{sect:Bayes}, we explain the Bayesian approach used in SPInS. In Sect. \ref{sect:Vstars}, we present some results obtained with SPInS for a set of fictitious stars with particularly noticeable locations in the H--R diagram.   In Sect. \ref{sect:Casa}, we compare results obtained by SPInS with those derived by \citet{2011A&A...530A.138C}.  In Sect.~\ref{sect:Single}, we show inferences on properties of stars observed either in interferometry or in asteroseismology.  In Sect.~\ref{sect:Ensembles}, we use SPInS to study coeval stars belonging to a binary system and an open cluster. Finally, we draw some conclusions in Sect. \ref{sect:Conclusion}.
	
	\section{Description of the SPInS code}
	\label{sect:Bayes}  
	
	\subsection{Overview}
	
	SPInS uses a Bayesian approach to find the PDF of the stellar parameters from a set of observational constraints.  At the heart of the code is a MCMC solver based on the Python EMCEE package \citep{2013PASP..125..306F} coupled to interpolation within a pre-computed grid of stellar models.  This allows SPInS to produce a sample of interpolated models representative of the underlying posterior probability distribution.  We recall that the posterior probability distribution can be obtained from the priors and the likelihood via Bayes' theorem:
	\begin{equation}
		P(\theta|O) = \frac{P(O|\theta) P(\theta)}{P(O)},
		\label{eq:Bayes_theorem}
	\end{equation}
	where $O$ are the observational constraints and $\theta$ the model parameters.  In other words, this theorem provides a way of calculating the probability distribution for model parameters given a set of observational constraints, as represented by the likelihood function $P(O|\theta)$, and priors $P(\theta)$.  The next two sections will deal with these two terms in more detail.
	
	\subsection{The priors}
	
	The priors represent our a priori knowledge of how the model parameters should behave.  For instance, one expects a higher number of low-mass stars than high-mass stars, and this can be expressed as a prior on the mass with a higher probability at low mass values.  As described in the following subsections, the priors will apply to the following stellar parameters given that we will be working with BaSTI stellar model grids \citep{2004ApJ...612..168P,2006ApJ...642..797P}: mass, age, and metallicity.  Of course, the choice of model parameters that are included in the priors depends on the parameters that describe the grid being used with SPInS.
	
	\subsubsection{Initial mass function}
	\label{sect:imf}
	
	The {initial mass function (IMF)} was first introduced by \citet{1955ApJ...121..161S}. It provides a convenient way of parametrising the relative numbers of stars as a function of their mass in a stellar sample \citep[see e.g. the review by ][]{2010ARA&A..48..339B}.
	
	The number $dN(m)$ of stars formed in the mass interval $[m, m + dm]$ reads $dN(m) = \xi(m) dm$ where $\xi(m)$ is the IMF. SPInS can handle two forms of the IMF: a one-slope version that reads
	\begin{eqnarray}
	\xi(m) \propto                                  %=k\times
	\begin{array}{ll}
	\left(\frac{m }{m_\mathrm{H}}\right)^{-\alpha}& \mbox{\ for\ } m_\mathrm{0} \ < m/M_\odot\le\ m_\mathrm{max},\\
	\end{array}
	\label{eq:priori_imf1}
	\end{eqnarray}
	and a two-slopes version,
		\begin{eqnarray}
\xi(m) \propto %=k\times
\left\{\begin{array}{ll}
\left(\frac{m }{m_\mathrm{H}}\right)^{-\alpha_1}& \mbox{\ for\ } m_\mathrm{H} \ < m/M_\odot\le\ m_\mathrm{0},\\

\left(\frac{m_0 }{m_\mathrm{H}}\right)^{-\alpha_1}\left(\frac{m }{m_0}\right)^{-\alpha_2}& \mbox{\ for\ } m_\mathrm{0} \ < m/M_\odot\le\ m_\mathrm{max}.\\		
\end{array}\right.
\label{eq:priori_imf}
\end{eqnarray}    
	The one parameter version (Eq. \ref{eq:priori_imf1}) is related to the IMF introduced by \citet{1955ApJ...121..161S} with the following parameter,
	\begin{eqnarray}
		\begin{array}{ll}
			\alpha=2.35  & \mbox{\ for\ } m_\mathrm{0}=0.40 < m/M_\odot \le m_\mathrm{max}=10.\\
		\end{array}
		\label{eq:priori_slopeimf1}
	\end{eqnarray}    			
	The two-parameters version (Eq. \ref{eq:priori_imf}) can be used to implement the canonical IMF from \citet[][section 9.1]{2013pss5.book..115K} which is suitable for stars in the solar neighbourhood. In that case, 	
	\begin{eqnarray}
	\left\{ 
	\begin{array}{ll}
	  \!\alpha_1=1.30\pm 0.30 &\!\!\!\! \mbox{\ for\ } m_\mathrm{H}=0.07 < m/M_\odot \le m_\mathrm{0}=0.50,\!\!\!\! \\
      \!\alpha_2=2.30\pm 0.36 &\!\!\!\! \mbox{\ for\ } m_\mathrm{0}=0.50 < m/M_\odot \le m_\mathrm{max}=150.\!\!\!\!\\  \end{array}
   \right.
	\label{eq:priori_slopeimf2}
  \end{eqnarray} 		
	Any other form of the IMF can easily be added to the SPInS program.
	
	\subsubsection{Stellar formation rate} 
	
	We restrict our working age domain to an upper limit of $13.8$ Gyr, that is roughly the age of the Universe, and we used the following uniform truncated {stellar formation rate (SFR)}:
	
	\begin{eqnarray}
		\lambda(\tau)=\left\{\begin{array}{rl}
			1 & \mbox{\ for\ }\tau _1=0\ \le\tau\le\ \tau _2\ \mbox{=\ 13.8\ Gyr },\\
			0 & \mbox{\ elsewhere.}\end{array}\right .
		\label{eq:priori_sfr}
	\end{eqnarray}
	This translates into a prior on the age of a star.
	
	\subsubsection{Metallicity distribution function} 
	
	We assume that the metallicity measurements are or will be available for the stars we want to age-date. Therefore, we do not introduce an a priori assumption on what their metallicity should be \citep[see the discussion in][]{2005A&A...436..127J}. Thus, by default, we adopt a flat prior on the metallicity [M/H] distribution function (MDF). 
	
	However, any prior on the MDF can be introduced into SPInS. As an example, in Sect.~\ref{sect:Casa}, we introduced  the prior adopted by \citet{2011A&A...530A.138C} and given in their Appendix A to correct for metallicity biases found in the Geneva-Copenhagen Survey.
	
	\subsection{The likelihood function}
	
	The likelihood function is used to introduce observational constraints.  Typically, these 	include constraints on classic observables, such as the luminosity, effective temperature, and metallicity.  However, as will be shown in the following, constraints on other observables may be used, such as the absolute magnitude in any photometric band,  colour indices, asteroseismic indices, radius, and whatever parameters are available with the grid of models being used with SPInS (as described in Sect.~\ref{sect:models}).  {These} constraints take on the form of probability distributions on the value of the parameter.  This leads to the following formulation for the likelihood function:
	\begin{equation}
		P(O|\theta) = \prod_i P_i(O_i  | \theta),
	\end{equation}
	where the $P_i$ represent the probability distributions on individual parameters resulting from the observational constraints, and $O_i$ the values of those parameters obtained for  a given set of model parameters $\theta$.  The probability distributions $P_i$ are typically normal distributions although other options are available with SPInS.
	
	\subsection{Variable changes}
	
	One of the features of SPInS is to allow variable changes.  For instance, one may have observational constraints on $\sqrt{L}$ rather than $L$ or may have a prior on $\log_{10} M$ rather than $M$.  SPInS allows such variable changes for a handful of elementary functions.  Of course, such changes need to be taken into account in a self-consistent way. In other words, the underlying probability distribution should not be altered.  Accordingly, variable changes on observed parameters are treated differently than those on model parameters.  To understand this, we recall the relationship between probability functions after a change of variables:
	\begin{equation}
		P_X(x) = P_{Y}\left(y(x)\right)\left|\frac{\mathrm{d}y}{\mathrm{d}x}\right|.
		\label{eq:proba_var_change}
	\end{equation}
	We then introduce this relation into Bayes' theorem and assume for simplicity that there is a single observed parameter and a single model parameter.  We then assume the prior and likelihood function apply to $f(\theta)$ and $g(O)$, respectively, instead of $\theta$ and $O$.  This leads to:
	\begin{eqnarray}
		P(\theta|O) &=&
		\frac{P(g(O)|\theta)\left|\frac{\mathrm{d}g(O)}{\mathrm{d}O}\right| 
			P(f(\theta))\left|\frac{\mathrm{d}f(\theta)}{\mathrm{d}\theta}\right|}
		{P(g(O))\left|\frac{\mathrm{d}g(O)}{\mathrm{d}O}\right|} \nonumber \\
		&=& \frac{P(g(O)|\theta)P(f(\theta))}{P(g(O))}
		\left|\frac{\mathrm{d}f(\theta)}{\mathrm{d}\theta}\right|.
		\label{eq:proba_param}
	\end{eqnarray}
	As can be seen, a change of variables on an observed constraint does not lead to any modification to the way the probability is calculated because the corrective terms cancel out.  In contrast, applying a prior to a different variable than the one used in the grid requires multiplying by the term $\left|\frac{\mathrm{d}f(\theta)}{\mathrm{d}\theta}\right|$.  SPInS accordingly takes this term into account using analytic derivatives of the elementary functions used in the variable change.
	
	\subsection{Grids of stellar models}
	\label{sect:models}
	
	SPInS can easily include any set of evolutionary tracks or isochrones available in the literature or calculated by the user. In this work, we used the BaSTI stellar evolutionary tracks available at \url{http://albione.oa-teramo.inaf.it/index.html} and described in \citet{2004ApJ...612..168P,2006ApJ...642..797P}. We chose to use these data rather than more recent ones in order to make comparisons with previous works. 
	
	In the BaSTI database, many sets of stellar tracks, all in the mass range $M\in [0.5M_\odot, 10M_\odot]$, are available. These models are well-suited to age-dating stars of different kinds: they cover evolutionary stages running from the zero-age main sequence (we do not use here the additional pre-main sequence grid provided for a narrower range of mass) to advanced stages, including the red-giant and horizontal branches, and a  metal abundance range $Z\in[0.0001, 0.04]$, where $Z$ is expressed in mass fraction. This interval of $Z$-values corresponds to number abundances of metals relative to hydrogen $\mathrm{[M/H]}\in[-2.27,+0.40]$, where $\mathrm{[M/H]}=\log (Z/X)-\log (Z/X)_\odot$ and $X$ is the hydrogen mass fraction. The value of $(Z/X)_\odot$ depends on the solar mixture under consideration. GN93's solar mixture \citep{1993oee..conf...15g} has a value $(Z/X)_\odot=0.0245$. 
	
	On the BaSTI website, the following grids are available:
	
	\begin{itemize}
	\item 
		\textsl{Canonical grid:} it corresponds to standard stellar models that do not include gravitational settling, radiative accelerations, convective overshooting, rotational mixing, but otherwise are based on recent physics, as detailed in \citet{2004ApJ...612..168P}. The models are based on GN93's solar mixture.
	\item 
		\textsl{Non-canonical grid:} the difference with the canonical grid is that models in this grid account for  core convective overshooting during the H-burning phase which may have a non-negligible impact on age. As described in \citet{2004ApJ...612..168P}, in these models, convective mixing is extended  over $0.2$ pressure scale-heights above the Schwarzschild's core for a stellar mass higher than $1.7\ M_\odot$, no overshooting is considered for a mass lower than $1.1\ M_\odot$, and a linear variation is assumed in-between.
	\item  
		\textsl{$\alpha$-enhanced model grids} (both canonical and non-canonical): their element mixture corresponds to a metal distribution typical of the Galactic halo and bulge stars, with $[\alpha/\mathrm{Fe]}=+0.4$, where $[\alpha/\mathrm{Fe]}$ is the decimal logarithm of the ratio of the number abundance of $\alpha$-elements (i.e. formed by $\alpha$-capture thermonuclear reactions) with respect to iron and referenced to the solar value ($[\alpha/\mathrm{Fe]}_\odot=0.0$). These grids are described in \citet{2006ApJ...642..797P}.
\end{itemize}
	
	For each evolutionary track in the BaSTI {database}, the variation of the luminosity, effective temperature, absolute $M_b$ magnitude, and colour indices  are provided as a function of the age and mass of the model star for several photometric systems. We here use the tracks given in the Johnson-Cousins system which provide $M_V$, and $(B-V),  (U-B), (V-I), (V-R),  (V-J), (V-K) ,  (V-L), (H-K)$ colours, but other photometric systems are available in the BaSTI database. All these quantities can be used indifferently in SPInS.
	
	In addition, we considered four quantities that can be inferred from BaSTI models straightforwardly: the photospheric radius $R_\star$ calculated from Stefan Boltzmann's law, the surface gravity
	\begin{eqnarray}
		\label{eq:gravity}
		{g= \frac{G M_\star}{R_\star^{2}},}
	\end{eqnarray}
	(or its decimal logarithm $\log g$),  and the frequency at maximum amplitude $\nu_\mathrm{max, sc}$ and the mean large frequency separation of pressure modes $\langle \Delta \nu \rangle_\mathrm{sc}$ expressed in asteroseismic scaling relations. The latter read,
	\begin{eqnarray}
		\label{eq:numax}
		{\frac{\nu_\mathrm{max, sc}}{\nu_\mathrm{max,\odot}} {=}\left(\frac{M_\star}{M_\odot}\right)\left(\frac{T_\mathrm{eff,\,\star}}{T_{\mathrm{eff},\,\odot}}\right)^{-1/2} \left(\frac{R_\star}{R_\odot}\right)^{-2},}
		%     {\langle \Delta \nu \rangle_\mathrm{sc}/\langle \Delta \nu \rangle_\odot{=}(M/M_\odot)^{1/2} (R/R_\odot)^{-3/2}}, 
	\end{eqnarray}
	and
	\begin{eqnarray}
		\label{eq:Dnu}
		%     {\nu_\mathrm{max, sc}/\nu_\mathrm{max,\odot} {=}(M/M_\odot)(T_\mathrm{eff}/5777)^{-1/2} (R/R_\odot)^{-2},}\\
		{\frac{\langle \Delta \nu \rangle_\mathrm{sc}}{\langle \Delta \nu \rangle_\odot}{=}\left(\frac{M_\star}{M_\odot}\right)^{1/2} \left(\frac{R_\star}{R_\odot}\right)^{-3/2}}, 
	\end{eqnarray}
	and are explained in \citet[][]{1994ApJ...427.1013B, 1995A&A...293...87K,2011A&A...530A.142B}. In Eqs. \ref{eq:gravity}, \ref{eq:numax}, and  \ref{eq:Dnu}, $G$ is the gravitational constant,  $M_\star$ the mass of the star, $T_{\mathrm{eff},\,\odot} = 5777$ K, the subscript `$\mathrm{sc}$' stands for scaling, and $\nu_\mathrm{max,\odot}=3090\ \mu$Hz and $\langle \Delta \nu \rangle_\odot=135.1\ \mu$Hz are the solar values in \citet{2011ApJ...743..143H}. In the following, $\nu_\mathrm{max, sc}$ and $\langle \Delta \nu \rangle_\mathrm{sc}$ will be referred to as seismic indices.
	
	\subsection{Interpolation in the grids}
	
	As was the case for the AIMS code, SPInS uses a two-step process for interpolation.  This then allows the MCMC algorithm to randomly select any point within the relevant parameter space.  The first part of the interpolation concerns interpolation between evolutionary tracks.  The second part concerns interpolation along the tracks, that is as a function of age.  These are described in the following subsections.
	
	\subsubsection{Interpolation between evolutionary tracks}
	
	Interpolation between evolutionary tracks amounts to interpolating in the parameter space defined by the grid parameters, excluding age.  In this parameter space, each track corresponds to a single point.  As a first step, a Delaunay tessellation is carried out for this set of points via the Qhull package\footnote{\url{http://www.qhull.org/}} \citep{Qhull} as implemented in SciPy\footnote{\url{https://www.scipy.org/}}.  As a result, the parameter space is subdivided into a set of simplices (i.e. triangles in two dimensions, tetrahedra in three dimensions, etc.). Then, for any point within the convex hull of the tessellation, SPInS searches for the simplex which contains the point and carries out a linear barycentric interpolation on the simplex.  The advantage of such an approach is that the grid of stellar models can be completely unstructured, thus providing SPInS with a greater degree of flexibility.  Furthermore, fewer tracks are linearly combined during the interpolation process thus potentially saving computation time compared to multilinear interpolation in Cartesian grids of the same number of dimensions.
	
	\subsubsection{Interpolation along evolutionary tracks}
	
	The second part of the interpolation focuses on age interpolation along evolutionary tracks.  Interpolation along a track is achieved by simple linear interpolation between adjacent points thus leading to piecewise affine functions for the various stellar parameters as a function of age.  What is more difficult is combining age interpolation with interpolation between tracks.  As opposed to AIMS, SPInS uses two variables for the age: the physical age and a dimensionless age parameter.  The purpose of the age parameter is to provide equivalent evolutionary stages on different tracks for the same value of this parameter.  This then allows SPInS to combine models at the same evolutionary stage when interpolating between tracks thus improving the accuracy of the interpolation.  Nonetheless, from the point of view of the MCMC algorithm, it is the physical age which is relevant, that is to say the MCMC algorithm will sample the physical age (thus bypassing the need for a corrective term as in Eq.~\ref{eq:proba_var_change}).  This is particularly important when fitting multiple coeval stars, that is with a common {physical} age.  Hence, SPInS is constantly going back and forth between these two age variables.
	
\begin{figure}[ht]
		\begin{center}
			\includegraphics[width=\linewidth]{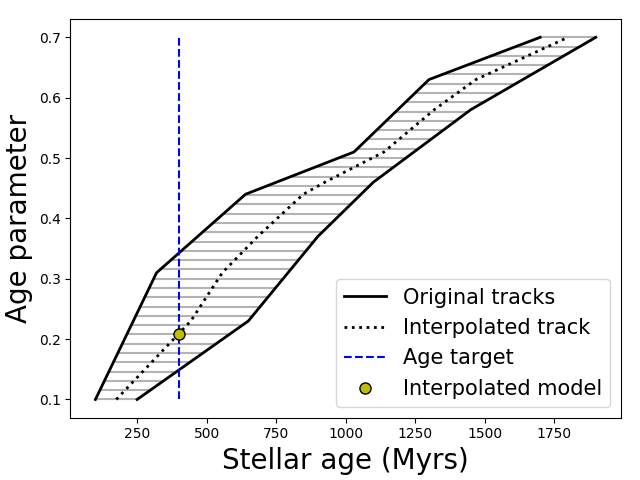}
			\caption{Schematic plot illustrating how age interpolation works in SPInS.  The two solid lines correspond to two neighbouring stellar evolutionary tracks which are involved in the interpolation.  The horizontal hatch marks indicate {that} the interpolation takes place horizontally (i.e. models with the same age parameter rather than physical age are linearly combined).  The dotted line shows the interpolated track.  The vertical blue dashed line corresponds to the target age and the yellow dot to the interpolated model.}
			\label{fig:age_interpolation}
		\end{center}
	\end{figure}

	Sampling as a function of physical age while interpolating in terms of the age parameter is not straightforward as illustrated in Fig.~\ref{fig:age_interpolation}.  In the plot, the interpolated track is halfway between the two original tracks for fixed values of the age parameter (although, SPInS can, of course, also interpolate using other interpolation coefficients). The same is not true for fixed stellar ages: the interpolated track is not halfway between the original tracks for fixed stellar ages, as can be seen for instance with the vertical dashed line at the target age.  Hence, one cannot simply find the age parameters on the original tracks for the target stellar age and interpolate between these to obtain the age parameter of the interpolated model.  One solution would be to interpolate the entire track and then search for the age parameter directly on it.  This is, however, not the most efficient approach computationally as most of the track is not needed and would probably considerably slow down SPInS given that this operation would need to be performed for each set of parameters tested by the MCMC algorithm.  The solution implemented in SPInS consists in a dichotomic search as a function of the age parameter combined with a direct resolution once the interval is small enough to only contain a single affine section of the interpolated track.  For the sake of efficiency, this part is written in Fortran.
	
	\subsection{Fitting multiple stars}
	
	As explained earlier, SPInS can simultaneously fit multiple coeval stars, such as what is expected in binary systems or stellar clusters.  Accordingly, a set of model parameters is obtained for each star, with however, the possibility of imposing common parameters such as age and metal content. Individual likelihood functions are defined for each star, whereas the same set of priors is applied to the model parameters for each star.  For common parameters, the prior is only applied once, unless the user specifically configures SPInS to apply it to each star (which amounts to raising the prior to the power $n_{\mathrm{stars}}$, where $n_{\mathrm{stars}}$ is the number of stars). Hence, the overall posterior probability is obtained as the product of the likelihood functions and priors applied to the parameters of each star apart from those of the common parameters which are only applied once.  Finally, for stellar samples sharing the same age, an isochrone file may be produced covering the whole mass interval spanned by the stellar models used by SPInS. Fitting multiple stars is advantageous as it can lead to tighter constraints on common parameters \citep[e.g.][]{2005A&A...436..127J}.
	
	\subsection{Typical calculation times}
	
	Computation times depend on a number of factors such as the number of stars being fitted, $n_{\mathrm{stars}}$, and the number of dimensions of the grid (excluding age), $n_{\mathrm{dim}}$, as well as on various MCMC parameters such as the number of iterations (both burn-in and production), $n_{\mathrm{iter}}$, the number of walkers, $n_{\mathrm{walk}}$, and the number of temperatures if applying parallel tempering, $n_{\mathrm{temp}}$.  Typical computation times for individual stars, $(n_{\mathrm{stars}},\, n_{\mathrm{dim}},\, n_{\mathrm{iter}},\, n_{\mathrm{walk}},\, n_{\mathrm{temp}}) = (1, 2, 400, 250, 10)$, is of the order of 1 min when using four processes on a Core i7 CPU.  When fitting 92 stars simultaneously from the Hyades cluster using age and metallicity as common parameters, $(n_{\mathrm{stars}},\, n_{\mathrm{dim}},\, n_{\mathrm{iter}},\, n_{\mathrm{walk}},\, n_{\mathrm{temp}}) = (92, 2, 600, 250, 10)$, the computation time was around 1.5 hours.  However, convergence is slower in such a situation given the higher number of dimensions from the point of view of the MCMC algorithm.  Hence, 20\,000 burn-in plus 200 production iterations were needed, thus leading to a computation time of roughly 75 to 150 hours using four processors (although this was carried out on a slightly slower processor).
	
	%--------------------------------------------------------------------
	
	\section{Parameter inference for a set of fictitious stars}
	\label{sect:Vstars} 
	
	%-----------------------------------------------
	\begin{table}
		\centering
		\caption{Set of fictitious stars to be characterised by SPInS. Here, inputs to SPInS are $\log(L/L_\odot)$, $\log(T_\mathrm{eff})$, and $\mathrm{[M/H]}$. For all stars, we adopted a solar metallicity $\mathrm{[M/H]}=0.00 \pm 0.05$ and the following  uncertainties $\sigma_{\log(L/L_\odot)}=0.04$ and  $\sigma_{T_\mathrm{eff}}=100$ K. The mass and age inferred by SPInS and their error bars are listed in Cols. 4 and 5.}
		\begin{tabular}{lcccc}
			\hline \hline
			star & $\log(\nicefrac{L}{L_\odot})$ & $\log(T_\mathrm{eff})$ & A (Gyr) &$\nicefrac{M}{M_\odot}$  \\
			\hline
           SF$ 1$ & $  0.00$ & $3.76$ & $  7.12\pm  3.72$ & $  0.96\pm  0.05$\\
           SF$ 2$ & $  0.25$ & $3.80$ & $  1.76\pm  1.30$ & $  1.16\pm  0.04$\\
           SF$ 3$ & $  0.25$ & $3.77$ & $  8.30\pm  2.20$ & $  1.02\pm  0.04$\\
           SF$ 4$ & $  0.50$ & $3.80$ & $  3.20\pm  0.85$ & $  1.23\pm  0.04$\\
           SF$ 5$ & $  0.50$ & $3.75$ & $  7.54\pm  0.92$ & $  1.10\pm  0.04$\\
           SF$ 6$ & $  0.50$ & $3.70$ & $  9.34\pm  2.14$ & $  1.06\pm  0.06$\\
          SF$ 7$ & $  0.83$ & $3.80$ & $  2.49\pm  0.44$ & $  1.45\pm  0.06$\\
          SF$ 8$ & $  1.00$ & $3.90$ & $  0.50\pm  0.18$ & $  1.70\pm  0.04$\\
          SF$ 9$ & $  1.00$ & $3.80$ & $  2.00\pm  0.24$ & $  1.59\pm  0.06$\\
          SF$10$ & $  1.00$ & $3.70$ & $  4.80\pm  2.33$ & $  1.31\pm  0.15$\\
          SF$11$ & $  1.50$ & $4.00$ & $  0.15\pm  0.06$ & $  2.29\pm  0.05$\\
          SF$12$ & $  1.50$ & $3.90$ & $  0.74\pm  0.04$ & $  2.12\pm  0.05$\\
          SF$13$ & $  1.50$ & $3.70$ & $  1.82\pm  1.05$ & $  1.79\pm  0.27$\\
          SF$14$ & $  2.00$ & $4.10$ & $  0.04\pm  0.02$ & $  3.14\pm  0.06$\\
         SF$15$ & $  2.00$ & $4.00$ & $  0.33\pm  0.02$ & $  2.82\pm  0.06$\\
         SF$16$ & $  2.00$ & $3.70$ & $  0.53\pm  0.10$ & $  2.77\pm  0.13$\\
         SF$17$ & $  3.00$ & $4.25$ & $  0.02\pm  0.01$ & $  5.63\pm  0.09$\\
         SF$18$ & $  3.00$ & $4.15$ & $  0.08\pm  0.00$ & $  4.97\pm  0.11$\\
        SF$19$ & $  3.00$ & $3.90$ & $  0.11\pm  0.01$ & $  4.70\pm  0.13$\\
         SF$20$ & $  3.00$ & $3.70$ & $  0.13\pm  0.01$ & $  4.63\pm  0.13$\\
			\hline
		\end{tabular}
		\label{tab:fictitious}
	\end{table}
	%--------------------------------------------------------------------------------------------------	
	
	\begin{figure}[ht]
		\begin{center}
			\resizebox{\hsize}{!}
			{\includegraphics{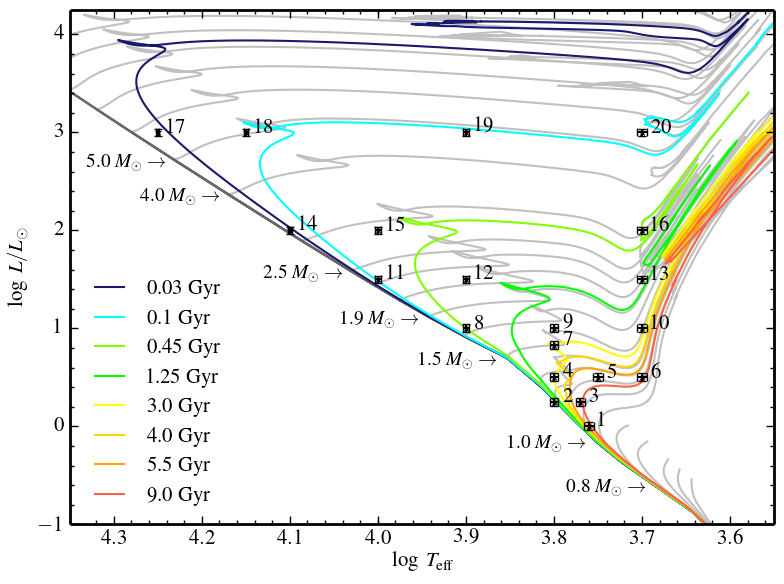}}
			\caption{Set of fictitious stars in the H--R diagram. Each star is labelled by its number in Table~\ref{tab:fictitious}. BaSTI stellar evolution tracks, some of which are labelled by their initial mass, are shown as well as isochrones, the ages of which are indicated in the legend starting from the youngest.}
			\label{fig:SF_HR}
		\end{center}
	\end{figure}
	
	\begin{figure*}[ht]
		\begin{center}
			\resizebox{\hsize}{!}
			{\includegraphics{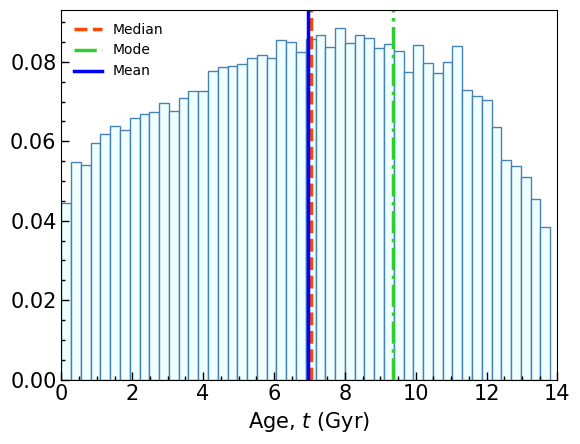}\includegraphics{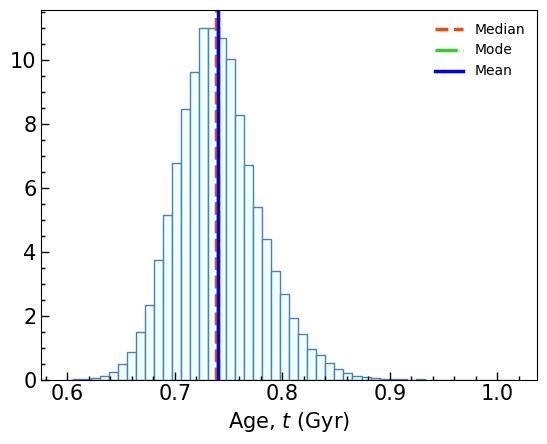}}
			\resizebox{\hsize}{!}{\includegraphics{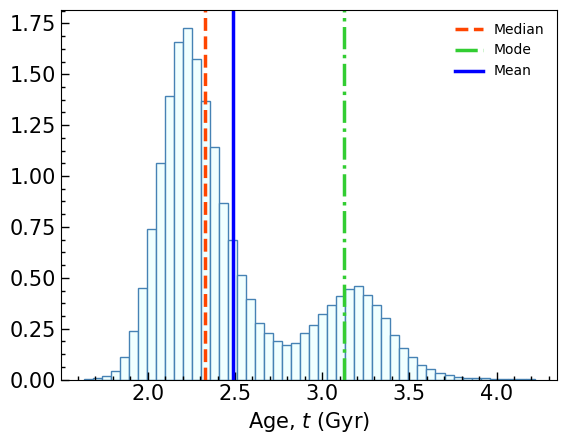}\includegraphics{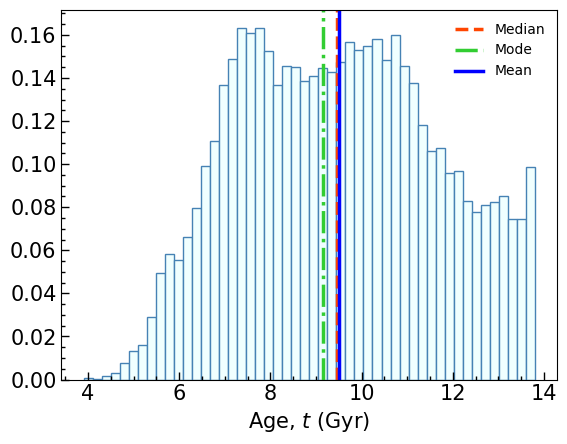}}
			\caption{Morphology of the age posterior probability distribution function (PDF) for different positions (i.e. evolutionary states) in the H--R diagram. The vertical lines indicate the values of the mean (blue continuous line), the mode (green dot-dashed line), and median (dashed red line). Top row, left panel: ill-determined (star SF1). Top row, right panel: one single peak (star SF12). Bottom row, left panel: age degeneracy leading to two peaks (star SF7). Bottom row, right panel: peculiar (star SF6). See discussion in the text.}
			
			\label{fig:SF_ppd}
		\end{center}
	\end{figure*}
	
	\begin{figure}[ht]
		\begin{center}
			\resizebox{\hsize}{!}
			{\includegraphics{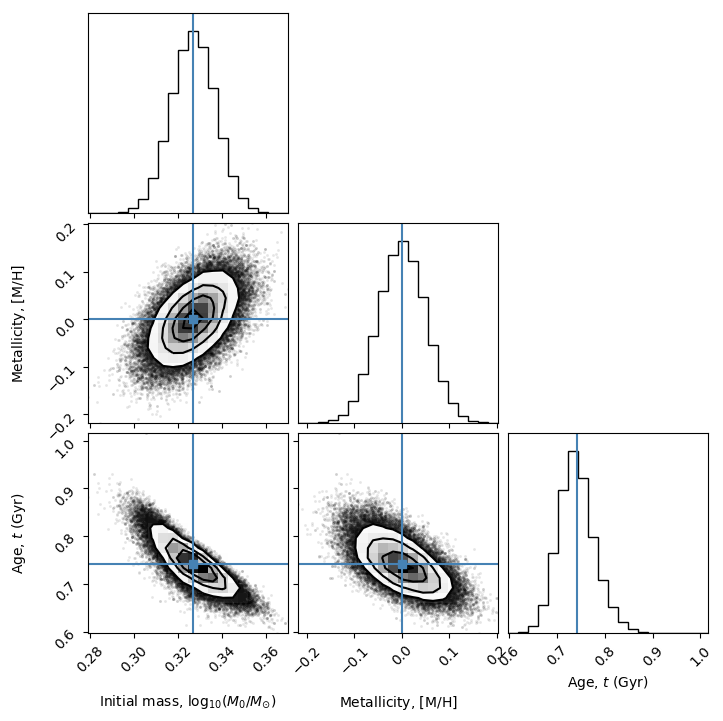}}
			\resizebox{\hsize}{!}
			{\includegraphics{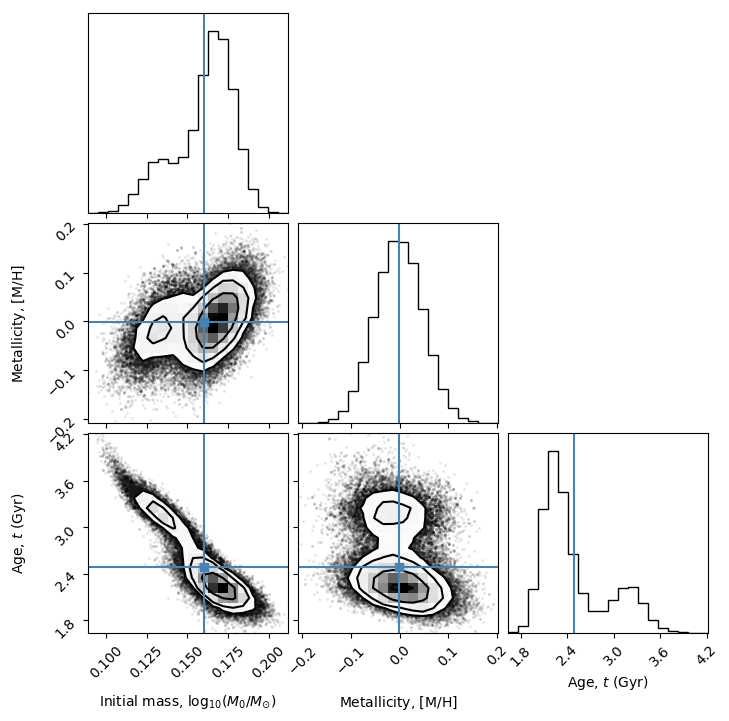}}
			\caption{Triangle plots with the distributions of fitted parameters including 2D correlations among parameters for a number of MCMC walkers. Mean values are indicated by blue continuous  lines. The point density is indicated by the greyscale of the distributions, with the darker, denser regions towards the maximum of the PDF. The contour lines indicate 0.5, 1.0, 1.5, and 2.0 sigma departures from the maximum. The top figure shows the results for star SF12, whereas the bottom figure corresponds to star SF7.} 
			\label{fig:triangle_SF}
		\end{center}
	\end{figure}
	
	As a case study, corresponding to the most common demand for age and mass inference on a survey-wide scale, we first focus on determining the properties of a small set of fictitious stars with solar metallicity and spread across the H--R diagram. The stars' properties (luminosity, effective temperature, metallicity, and their error bars)
	used as inputs to SPInS are listed and explained in Table~\ref{tab:fictitious} and its caption. The positions of the stars in the H--R diagram are shown in Fig.~\ref{fig:SF_HR}. We used the solar-scaled  non-canonical BaSTI stellar model grid (cf. section \ref{sect:models}), as well as the IMF from \citet{2013pss5.book..115K} given in Eqs. \ref{eq:priori_imf} and \ref{eq:priori_slopeimf2}, a uniform truncated SFR (Eq. \ref{eq:priori_sfr}), and a uniform MDF, as priors. For each star, the inferred age and mass are listed in Table~\ref{tab:fictitious}. Depending on the position of a star in the H--R diagram, the solution may be subject to an important degeneracy which is revealed in the posterior probability distribution function \citep[see the thorough discussions in, e.g.][]{2005A&A...436..127J, 2007ApJS..168..297T}. We show in Fig.~\ref{fig:SF_ppd} several PDFs showing a different typical morphology which we examine below.
	
	\begin{itemize}
		\item Firstly, low-mass stars on the main sequence (hereafter MS) lie in a region where the isochrones are crowded and thus can be fitted by practically any isochrone. This is the case of star SF1, close to the Sun, whose age is very ill-defined (see the PDF in the left panel of top row in Fig.~\ref{fig:SF_ppd}).
		
		\item Secondly, more massive stars, either on the MS or fully installed on the subgiant branch, have a rather well-defined age and their PDF shows a single peak. This is the case for star SF12 shown in the top row, right panel of Fig.~\ref{fig:SF_ppd}, but also for MS stars SF3, SF8, SF11, SF14, SF15, SF17, and  subgiants SF5 and SF19. For stars close to the zero-age MS such as star SF2, that is barely evolved stars, the one-peak PDF (not shown) is very asymmetric. It is  truncated close to age {`zero'} of the evolutionary tracks, meaning that for these stars, tracks including the pre-main sequence phases should be used. Moreover, since the region where star SF2 lies is still crowded with isochrones, its PDF  shows a very long tail towards high ages up to about 8 Gyr.
		
		\item Thirdly, for stars of mass $M\gtrsim 1.2 M_\odot$ which had a convective core during the MS and are now lying in the so-called hook region, in the vicinity of the end of the MS, several ages are possible. However, these ages are not equally probable because of the different amounts of time spent either before the red point of the evolutionary track (i.e. minimum of $T_\mathrm{eff}$ on the hook) or in the second contraction phase before the blue point (i.e. maximum of $T_\mathrm{eff}$ on the hook), or at the very beginning of the subgiant phase. This translates into a PDF generally showing two peaks, as can  be seen in the {bottom row, left panel} of Fig.~\ref{fig:SF_ppd} for star SF7, but the same behaviour is seen for stars SF4, SF9, and SF18. For star SF20, located close to the helium burning region where the star undergoes blue loops, the PDF also shows two peaks, one of them being very discreet. 
		
		\item Fourthly, stars lying close to the red giant branch either show a more or less well-defined peak in their PDF (such as stars SF13 and SF16) or a flattened PDF (such as star SF6 shown in the bottom row, right panel of Fig.~\ref{fig:SF_ppd} and star SF10).  
	\end{itemize}
	
	Several indicators of a parameter, for instance the age, can be used. In Fig.~\ref{fig:SF_ppd}, we show the median, the mean, and the posterior mode values given by SPInS for stars SF1, SF6, SF7, and SF12. The estimator of the mode is the maximum a posteriori (MAP). However, in the case of star SF7, the age PDF is multimodal, showing two maxima.  Figure~\ref{fig:SF_ppd} shows that the mode provided by SPInS is close to the second maximum when, intuitively, one would have taken the mode to be the value at the maximum of the higher peak. It is because the mode, as calculated by SPInS, corresponds to the parameter set yielding the maximum posterior probability given the observations (see Eq.~\ref{eq:Bayes_theorem}) obtained in the grid's parameter space (i.e. in the mass-age-metallicity space for this particular example).  In contrast,
	the histogram showing the age PDF is obtained after an integration (i.e. marginalisation) with respect to mass and metallicity.  For star SF7, the secondary peak in the age histogram corresponds to a higher but narrower peak (especially in terms of the variables mass and metallicity) in the original grid parameter space, thus explaining why it is lower after marginalisation.
	On the other hand, the mass is well-determined for all stars, with PDFs mostly presenting one or two peaks. 
	
	In addition, SPInS provides triangle plots showing the distributions of the fitted parameters and their correlations. Examples are shown in Fig.~\ref{fig:triangle_SF} for stars SF12 (top figure) and SF7 (bottom figure) where the distributions of age, mass, and metallicity are shown. More complete triangle plots (not shown here) showing other parameters (for instance the radius, surface gravity, seismic {indices}, etc.) can be plotted according to the user's choice. SPInS also provides a number of files and figures making the analysis of the intermediate and final results easy, including files listing the mean values of each estimated parameter, the median, the one- and two-sigma percentiles, the correlations among parameters, and the set of parameters corresponding to the mode.
	
	\section{Parameter determination for stars in the Geneva-Copenhagen Survey}
	\label{sect:Casa} 
	
	\begin{figure*}[htb]
		\begin{center}
			\resizebox{\hsize}{!}
			{\includegraphics{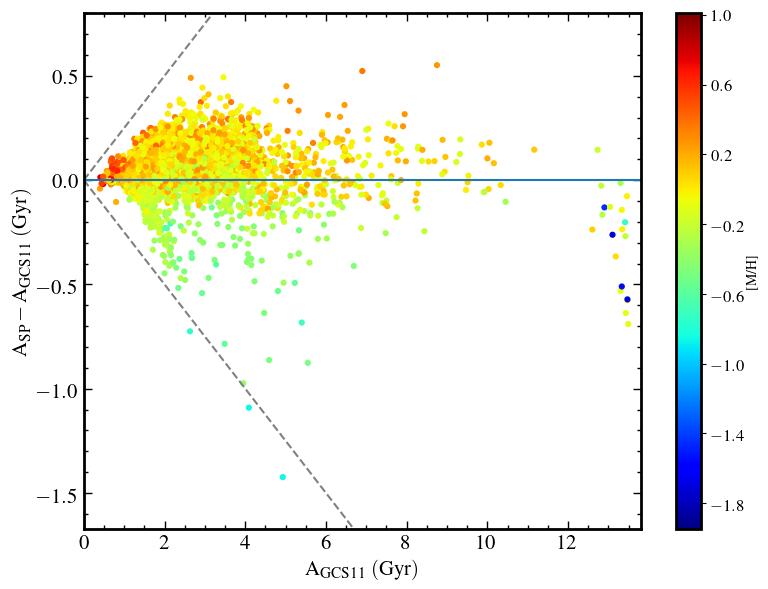}\includegraphics{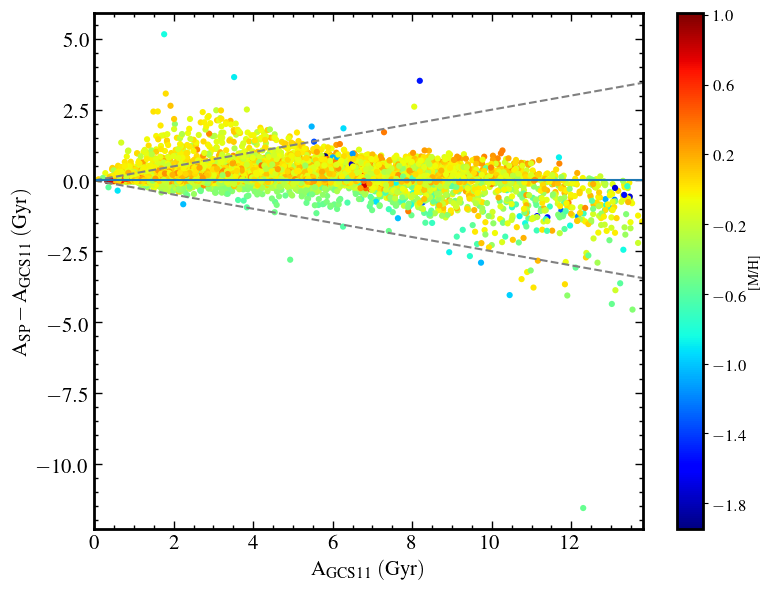}}
			\caption{Age residuals ($\mathrm{A_{SP} -A_{GCS11}}$) between the mean age value  delivered by SPInS ($\mathrm{A_{SP}}$) and the expected (i.e. also mean) age ($\mathrm{A_{GCS11}}$) given in the GCS11 \citep{2011A&A...530A.138C}. The horizontal grey line shows the locus of equal ages and the dashed lines are for ages differing by $\pm 25$ per cent. The colours distinguish stars according to their metallicity. Left figure:  5040  stars with good ages as defined by \citeauthor{2011A&A...530A.138C} (see text), with no filtering of the parallax error. Right figure: 10\,865 stars with determined ages, all stars having a parallax error lower than 10 per cent.}
			\label{fig:ages_GCS11}
		\end{center}
	\end{figure*}
	
	In this section, we aim at testing and validating the SPInS tool by characterising the stars of the Geneva-Copenhagen Survey (GCS). The GCS is a compilation of observational and stellar model-inferred properties of stars belonging to the solar neighbourhood. The first version of the GCS was presented and made public by \citet[][hereafter GCS04]{2004A&A...418..989N}. It provides a complete, magnitude-limited ($V<8.3$), and kinematically unbiased sample of 16\,682 nearby F and G dwarf stars. Many data can be found in the catalogue, of {which} the Hipparcos parallax, metallicity, effective temperature, and Johnson $V$-magnitude {are of interest here}. Later on, the data in the catalogue were assessed and refined by \citet[][hereafter GCS09]{2007A&A...475..519H, 2009A&A...501..941H}. Then, \citet[][hereafter GCS11]{2011A&A...530A.138C} improved the accuracy of the effective temperatures on the basis of the infrared flux method, and consequently improved the metallicity scale. They also provided a proxy for the $[\alpha/\mathrm{Fe}]$ ratio and reddening E(B-V). Each version of the GCS also provides the age and mass of the stars, and related uncertainties, derived by means of a Bayesian analysis.
	
	In order to compare the results of SPInS with those of  \citet{2011A&A...530A.138C}, we used SPInS to determine the ages and masses of stars in the GCS. We adopted, as far as possible, the assumptions made by these authors, namely:
	
	\begin{itemize}
		\item We used the following observational constraints for each star: logarithm of effective temperature $\log T_\mathrm{eff}$, absolute $V$-magnitude in the Johnson band $M_V$, and metallicity [M/H] (not [Fe/H]).
		
		\item We took the same source for stellar models, that is the solar-scaled canonical BaSTI grid described in Sect.~\ref{sect:models} taken from their website \citep{2004ApJ...612..168P}. However, the grid used by \citet{2011A&A...530A.138C} was specially prepared and is finer than the one available on the  website. Furthermore, in contrast with  \citeauthor{2011A&A...530A.138C}, we did not use the isochrones but the evolutionary tracks, which are the direct products of stellar evolution calculations.
		
		\item As was done in \citet{2011A&A...530A.138C}, we adopted stellar evolution models calculated for a solar-scaled mixture, but we re-scaled the metallicity to mimic the $\alpha$-element enrichment. Since \citet{2011A&A...530A.138C} do not explicitly give the re-scaling relation they adopted, we adopted the most commonly used relation derived by \citet{1993ApJ...414..580S}\footnote{In \citet{1993ApJ...414..580S}, the solar-scaled $Z$ is re-scaled as $Z_\alpha$ according to $Z_\alpha=Z\times(0.638\times 10^{\mathrm{[\alpha/Fe]}}+0.362)$}. However, we checked that only minor differences are obtained if the re-scaling of \citet{2004A&A...418..989N}\footnote{$\mathrm{[M/H]}_\alpha=\mathrm{[M/H]}+0.75\times\mathrm{[\alpha/Fe]}$}, applied in the GCS04, is used instead.
		
		\item To calculate the absolute magnitudes $M_V$ of each star, we used the Johnson $V$-magnitude provided in the GCS09 and the Hipparcos  parallax provided in the GCS11 \citep[i.e. the so-called new Hipparcos reduction from][]{2007A&A...474..653V}. We corrected the absolute magnitudes for the effects of extinction following \citet{1989ApJ...345..245C} with E(B-V) taken from the GCS11.
		
		\item We assumed a Gaussian distribution in $\log T_\mathrm{eff}$, $M_V$, and [M/H]. We, therefore, did not implement in SPInS the particular treatment of the magnitude distribution adopted by \citet{2011A&A...530A.138C} to take into account the skewness of the magnitude distribution that appears when the relative parallax error exceeds 10 per cent. Therefore, the present comparisons will not be valid for stars with high parallax {errors}.
		
		\item We adopted the same priors on the IMF, SFR, and MDF. More precisely, we took the IMF from \citet[][]{1955ApJ...121..161S} as given by   Eqs. \ref{eq:priori_imf1} and \ref{eq:priori_slopeimf1}, a uniform truncated SFR (Eq. \ref{eq:priori_sfr}), and the prescription of \citet{2011A&A...530A.138C} for the particular MDF of the stars in the GCS11 (see their Appendix A).
	\end{itemize}
	
	\subsection{Ages}
	
	We present in Fig.~\ref{fig:ages_GCS11} the {age residuals between} the ages obtained by SPInS (mean ages) and those of the GCS11 (referred to as `expected' age in their terminology but which also corresponds to the mean value). The ages of 14\,757 stars out of 16\,682 could be determined. For the remaining stars, either the parallax, effective temperature, or metallicity were unavailable.  In Fig.~\ref{fig:ages_GCS11} (left panel), we only retained the stars for which the age-dating is considered to be of good quality under the criteria of \citet{2011A&A...530A.138C}, that is the relative error on age is lower than 25 per cent and the absolute age error is lower than 1 Gyr. This represents a total of  5040 stars. The comparison is very satisfactory since the differences in ages between SPInS and the GCS11 mostly remain lower than 25 per cent. In Fig.~\ref{fig:ages_GCS11} (right panel), we considered all stars with a determined age and a parallax error lower than 10 per cent (to minimise the possibility of a skewness in the $V$-magnitude distribution, see the discussion in Sect. \ref{sect:Casa} above). Even if the differences between the SPInS and GCS11 ages are larger in that case, there are only 503 stars out of {10\,865} (that is less than 5 per cent) showing an age difference larger than 25 per cent. 
	
	\subsection{Masses}
	
	\begin{figure}[ht]
		\begin{center}
			\resizebox{\hsize}{!}
			{\includegraphics{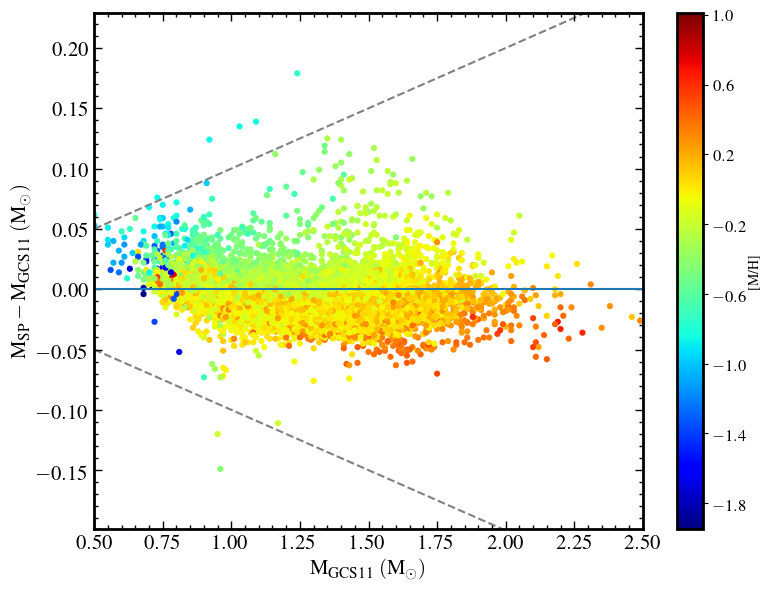}}
			\caption{Mass residuals ($\mathrm{M_{SP} -M_{GCS11}}$) between the mean mass value  ($\mathrm{M_{SP}}$) delivered by SPInS and the expected (i.e. also mean) mass ($\mathrm{M_{GCS11}}$) given in the GCS11  \citep{2011A&A...530A.138C}  for the 12\,704 stars with good masses as defined in the text, with no filtering on the parallax error. For ten stars, the masses  determined by SPInS and by \citeauthor{2011A&A...530A.138C}  differ by more than 10 per cent.  The horizontal grey line shows the locus of equal masses and the dashed lines are for masses differing by $\pm 10$ per cent. The colours distinguish stars according to their metallicity. }
			\label{fig:masses_GCS11}
		\end{center}
	\end{figure}
	
	In Fig.~\ref{fig:masses_GCS11}, we {show the mass residuals between} the masses (mean values) derived by SPInS and those in the GCS11 (expected values corresponding to mean ones). As for the ages, the masses of 14\,757 out of 16\,682 could be determined.  In the figure, we only retained stars for which we consider the mass to be of good quality, that is the relative error on mass is lower than 10 per cent. This represents a total of 12\,704 stars. The comparison is very satisfactory since the differences in mass between SPInS and GCS11 values mostly remain lower than 10 per cent. 
	The disparity of SPInS and GCS11 results is less important for the masses than for the ages because, for a given set of stellar evolutionary tracks, the mass degeneracy in the Hertzsprung-Russell diagram is less marked than the degeneracy affecting the ages, in particular at the end of the main sequence.
	
	\subsection{Radii, surface gravities, and mean seismic parameters}

	\begin{figure*}[ht]
		\begin{center}
			\resizebox{\hsize}{!}{\includegraphics{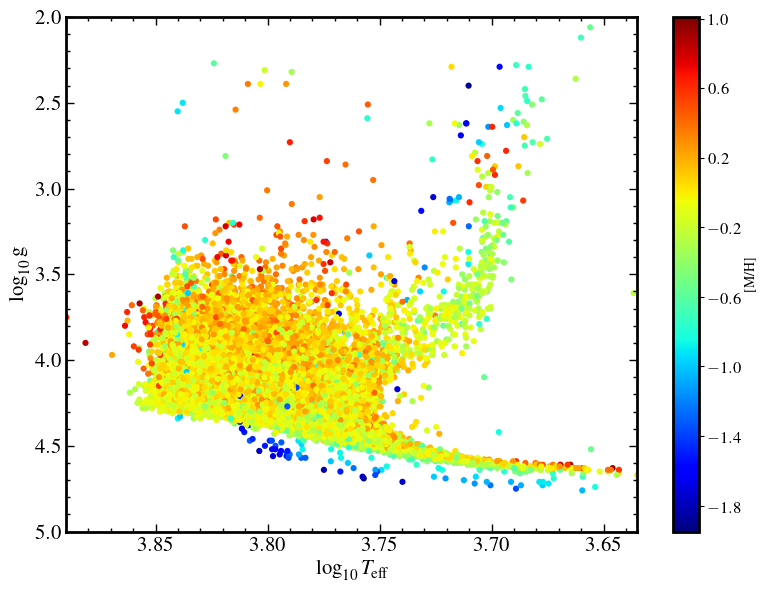}\includegraphics{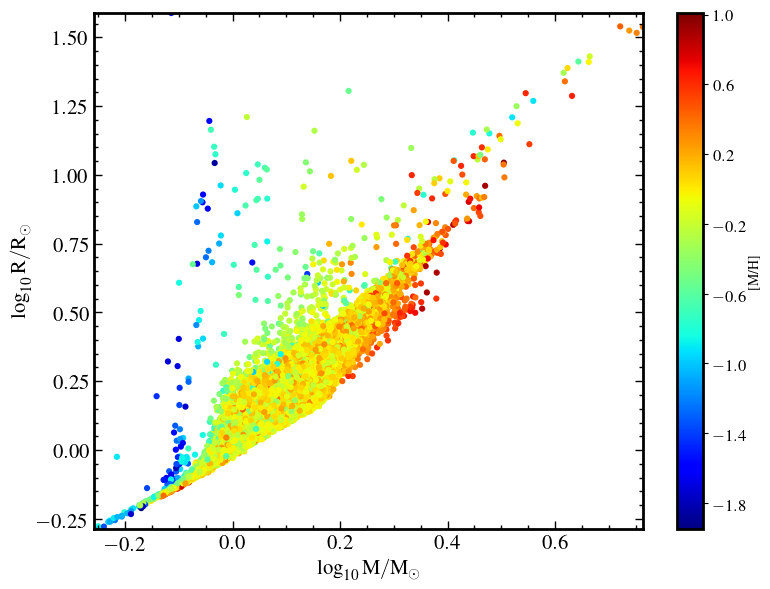}\includegraphics{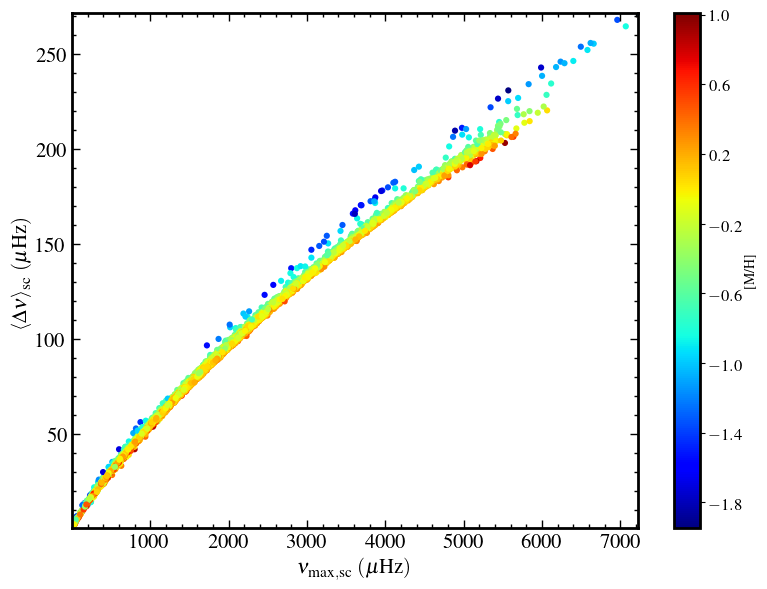}}
			\caption{Stellar parameters inferred with SPInS for about {14\,750} stars in the GCS11 catalogue. Left figure: Kiel diagram. Central figure: Mass-Radius relation. Right figure: Asteroseismic $\nu_\mathrm{max, sc}$--$\langle\Delta\nu\rangle_\mathrm{sc}$ diagram.}
			\label{fig:results_GCS11}
		\end{center}
	\end{figure*}	
	
	In addition to the ages and masses of the stars in the GCS11, SPInS provided several interesting stellar properties related to stellar evolutionary tracks or that can easily be derived from them. In particular, we show in Fig.~\ref{fig:results_GCS11} the Kiel diagram ($\log T_\mathrm{eff}-\log {g}$), the mass-radius relation, and the asteroseismic diagram ($\nu_\mathrm{max, sc}$--$\langle\Delta\nu\rangle_\mathrm{sc}$) based on seismic indices. The discussion of the results is beyond the scope of this paper, although some well-known trends can be highlighted, in particular the position of stars as a function of their metallicity in the Kiel diagram resulting from their different internal structures. It is worth pointing out that the combination of such diagrams, involving large stellar samples, can be very valuable when used in studies of Galactic Archaeology \citep[][]{2009A&A...503L..21M,2013MNRAS.429..423M} thus making SPInS a very interesting tool in this respect. 
	
	\section{Further applications of SPInS for single stars}
	\label{sect:Single}
	
	\subsection{Stars observed in interferometry}
	
	With interferometry, the angular diameters of stars can be measured which in turn gives a direct access to their radii, provided their distance is known. These measurements are therefore independent of stellar models (except for limb-darkening of the stellar disc which has to be corrected for, based on stellar model atmospheres). \citet{2016A&A...586A..94L} obtained the radii of $18$ stars (eleven of them being exoplanet hosts) from interferometry, together with their bolometric fluxes from photometry which allowed them to infer the effective temperatures. Starting from these data, metallicities taken in the literature, and model isochrones, they applied a Bayesian method with flat priors  to infer the mass and age of each star.
	
\begin{figure}[hb]
		\begin{center}
			\resizebox{1.0\hsize}{!}{\includegraphics{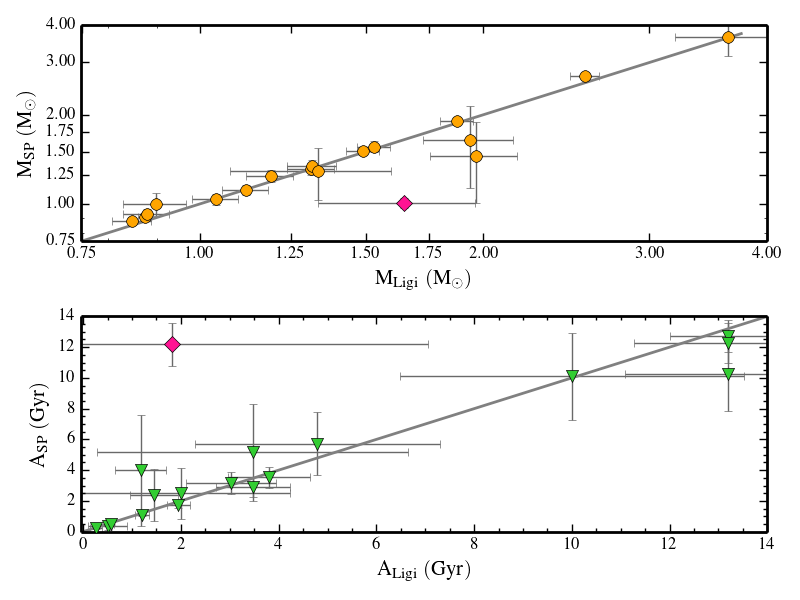}}
			\caption{Comparison of the {masses and ages} inferred by SPInS {with} the values obtained by \citet{2016A&A...586A..94L} for $18$ stars observed in interferometry. Top figure: Mass comparisons (orange circles). Bottom figure: Age comparisons (green triangles). The outlier star HD 167042 (see text) is shown with pink diamonds. The diagonal lines represent the one-to-one relation.}
			\label{fig:Ligi}
		\end{center}
	\end{figure}

	By using \citeauthor{2016A&A...586A..94L}'s radii, effective temperatures, and metallicities as input constraints {for} SPInS, we inferred the {masses and ages} of the stars. We used the solar-scaled non-canonical BaSTI grid including convective core overshooting and we assumed flat priors on mass and metallicity, but a uniform truncated prior on age (Eq. \ref{eq:priori_sfr}). In Fig.~\ref{fig:Ligi}, we compare SPInS masses and ages with {the values from} \citet{2016A&A...586A..94L}. Overall, the comparison is satisfactory, except for one   star, \object{HD 167042}. If we exclude this star, the mean mass (respectively age) difference is of 4 (respectively 19) per cent and maximum differences are of 15 (respectively 84) per cent. For HD 167042, shown with pink diamonds in Fig.~\ref{fig:Ligi}, SPInS's mass is much smaller than the one found by  \citet{2016A&A...586A..94L} while SPInS's age is much larger. Understanding the origin of the difference is beyond the scope of this paper. However, \citeauthor{2016A&A...586A..94L} pointed out that, for this star, their results are not consistent with the models. Moreover, we point out that it may {currently} be difficult to characterise this K1IV subgiant. Indeed, doubts remain about its effective temperature which is found to be $4547\pm49$ K when combining interferometry and photometry  \citep{2016A&A...586A..94L} 
	and $4983\pm10$ K {when} using high resolution spectroscopy \citep{2016A&A...588A..98M}. 
	
	{In the future, large samples of stars with angular radii measured by interferometry will be available. In particular, the CHARA/SPICA project\footnote{\url{https://lagrange.oca.eu/fr/spica-project-overview}}, based on the  new visible interferometric instrument CHARA/SPICA currently under design \citep{2018SPIE10701E..20M}, aims at constituting a homogeneous catalogue of about a thousand angular diameters of stars spanning the whole H--R diagram, including hosts of exoplanetary systems and stars observable in asteroseismology. SPInS will enable a rapid characterisation of the fundamental parameters of these stars (mass, age), thus opening the way to an in-depth analysis of their internal structure, planet characterisation, etc. }	
	
	\subsection{Solar-like oscillators}
	
	SPInS has not been designed for the purpose of delivering a precise asteroseismic diagnosis. To perform detailed asteroseismic analysis, one can use, for instance, the public AIMS tool described in \citet{2018ASSP...49..149L} and  \citet{2019MNRAS.484..771R}. However, when individual frequencies cannot be extracted from the pressure-mode oscillation spectrum, SPInS can give some characteristics of a star provided  the seismic {indices} $\nu_\mathrm{max, obs}$ or $\langle\Delta\nu\rangle_\mathrm{obs}$, or both have been estimated from observations. Indeed, SPInS can take $\nu_\mathrm{max, obs}$ and $\langle\Delta\nu\rangle_\mathrm{obs}$ as input constraints which, through the scaling relations (Eqs. \ref{eq:numax} and \ref{eq:Dnu}), provides hints on the stellar mass and radius, provided the effective temperature is known. In the following, we study two stellar samples (an artificial and a real one), for which SPInS inferences of mass, radius, and age based on $\nu_\mathrm{max, obs}$ and $\langle\Delta\nu\rangle_\mathrm{obs}$ can be compared with the results of careful inferences based on individual oscillation frequencies.
	
	\subsubsection{Artificial stars: \citet{2016A&A...592A..14R}'s hare and hounds sample}
	
	%-----------------------------------------------
	\begin{table*}
		\centering
		\caption{Simulated properties for the set of artificial stars of \citet{2016A&A...592A..14R}.  These properties are taken from their Table~1 except for $\langle\Delta \nu\rangle_\mathrm{obs}$ that we calculated as the least-square mean of the individual frequencies of radial modes given in Table A.1 of \citeauthor{2016A&A...592A..14R} (`obs' therefore corresponds to simulated results including a realistic error realisation). The `true' masses, radii, and ages (i.e. from the original stellar models) are provided in the last three columns.
		}
		\begin{tabular}{lcccccccc}
			\hline \hline
			star & $L_\mathrm{obs}$ & $T_\mathrm{eff, obs}$ &$\mathrm{[M/H]_{obs}}$  &$\nu_\mathrm{max, obs}$  &$\langle\Delta \nu\rangle_\mathrm{obs}$& $M$&  $R$& $A$\\
			& $(L_\odot)$ & (K) &  &$(\mu\mathrm{Hz})$  &$(\mu\mathrm{Hz})$ &$(M_\odot)$& $(R_\odot)$& (Gyr)\\
			\hline  
			Aardvark & $0.87  \pm 0.03$ &$5720  \pm  85$ & $  +0.02  \pm  0.09$ &   $3503  \pm  165$  &  $144.47  \pm  0.01$ & 1.00 &  0.959 &  3.058 \\
			Blofeld  & $2.02  \pm 0.06$ &$5808  \pm  85$ & $  +0.04  \pm  0.09$ &   $1750  \pm  100$  &  $ 94.14  \pm  0.01$ & 1.22 &  1.359 &  2.595 \\
			Coco     & $0.73  \pm 0.02$ &$5828  \pm  85$ & $  -0.74  \pm  0.09$ &   $3634  \pm  179$  &  $162.14  \pm  0.02$ & 0.78 &  0.815 &  9.616 \\
			Diva     & $2.14  \pm 0.06$ &$5893  \pm  85$ & $  +0.03  \pm  0.09$ &   $2059  \pm  101$  &  $ 95.77  \pm  0.01$ & 1.22 &  1.353 &  4.622 \\
			Elvis    & $1.22  \pm 0.04$ &$5900  \pm  85$ & $  +0.04  \pm  0.09$ &   $2493  \pm  127$  &  $119.96  \pm  0.01$ & 1.00 &  1.087 &  6.841 \\
			Felix    & $4.13  \pm 0.12$ &$6175  \pm  85$ & $  +0.06  \pm  0.09$ &   $1290  \pm   66$  &  $ 69.39  \pm  0.02$ & 1.33 &  1.719 &  2.921 \\
			George   & $4.31  \pm 0.13$ &$6253  \pm  85$ & $  -0.03  \pm  0.09$ &   $1311  \pm   67$  &  $ 70.25  \pm  0.04$ & 1.33 &  1.697 &  2.944 \\
			Henry    & $1.94  \pm 0.06$ &$6350  \pm  85$ & $  -0.35  \pm  0.09$ &   $2510  \pm  124$  &  $116.46  \pm  0.04$ & 1.10 &  1.138 &  2.055 \\
			Izzy     & $2.01  \pm 0.06$ &$6431  \pm  85$ & $  -0.34  \pm  0.09$ &   $2319  \pm  124$  &  $115.85  \pm  0.03$ & 1.10 &  1.141 &  2.113 \\
			Jam      & $3.65  \pm 0.11$ &$6503  \pm  85$ & $  +0.09  \pm  0.09$ &   $1758  \pm   89$  &  $ 86.54  \pm  0.06$ & 1.33 &  1.468 &  1.681  \\
			\hline
		\end{tabular}
		\label{tab:HHg}
	\end{table*}
	%-----------------------------------------------
	
	\begin{figure}[htb]
		\begin{center}
			\resizebox{1.0\hsize}{!}{\includegraphics{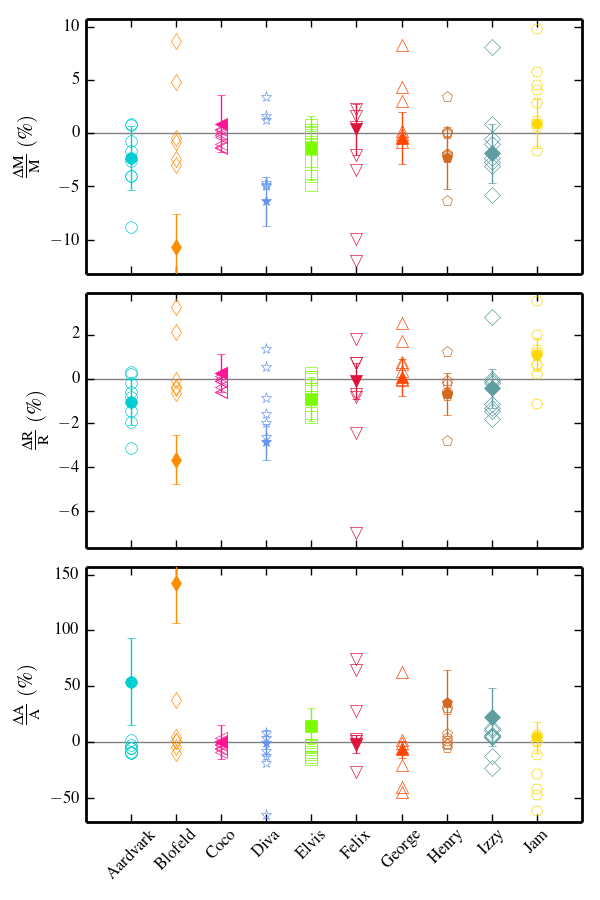}}
			\caption{{Comparison} of the masses, radii, and ages inferred by different techniques {with} the true values of the ten artificial stars of \citet{2016A&A...592A..14R}. Full symbols with error bars {correspond to} SPInS inferences with $\nu_\mathrm{max, obs}$ and $\langle\Delta\nu\rangle_\mathrm{obs}$ taken as seismic constraints. Open symbols are for the results of eight pipelines where a full seismic diagnosis based on individual frequencies has been performed \citep[see ][for details]{2016A&A...592A..14R}.}
			\label{fig:HH}
		\end{center}
	\end{figure}
	
	As a first case study, we consider ten artificial stars, built and studied  in the hare-and-hounds exercise of \citet{2016A&A...592A..14R}. To build each star, a stellar model was calculated for a given mass and age. Then, starting from this model, a hare group simulated observational quantities of an artificial star, that is its oscillation frequencies and classical parameters. Finally, the results were communicated to several hound teams who applied distinct optimisation methods to characterise the stars on the basis of these constraints. We list the properties of the ten stars in Table~\ref{tab:HHg}. Their positions in the H--R diagram are shown in Fig. 1 of \citet{2016A&A...592A..14R}.
	
	We applied SPInS to these stars taking their `observed'\footnote{Here, `observed' values correspond to simulated results including a realistic error realisation.} luminosity, effective temperature, metallicity, $\nu_\mathrm{max, obs}$, and $\langle\Delta\nu\rangle_\mathrm{obs}$ as input constraints. We used the solar-scaled non-canonical BaSTI grid including convective core overshooting and we took flat priors on mass, age, and metallicity.
	In Fig.\ref{fig:HH}, we show how the masses and radii inferred both by SPInS and by the teams that participated in the exercise of \citet{2016A&A...592A..14R} reproduce the true properties of the --artificial-- stars. 
	With SPInS, mean differences with the artificial stars are of  $2.8$ per cent on the predicted mass and of $1.1$ per cent on the radius. 
	The maximum differences are for Blofeld and to a lesser extent Diva. For Blofeld, the difference is of about $11$ per cent on mass and of $3.7$ per cent on radius. It is worth noting that both of these simulated stars have the same mass ($M_\mathrm{obs}=1.22 M_\odot$) and are both on the subgiant branch {but} have different chemical compositions. 
	Moreover, the BaSTI grid models \citep{2004ApJ...612..168P} have not been calculated with the same input physics and parameters as \citet{2016A&A...592A..14R}'s models. Indeed, \citeauthor{2016A&A...592A..14R}'s models of Diva and Blofeld were calculated with different amounts of convective core overshooting. Moreover, Blofeld  includes atomic diffusion, a different solar mixture, and  a truncated atmosphere. We also note that Diva is one of the least well-fitted stars in \citet{2016A&A...592A..14R}. 
	
	Overall, as can be seen in Fig.\ref{fig:HH}, except for the cases of Blofeld and Diva that, in all likelihood result from identified differences in stellar models, the masses and radii inferred by SPInS compare very well with those inferred from a thorough asteroseismic diagnosis based on individual oscillations frequencies.
	This confirms the power of the scaling relations to quite reasonably infer the mass and radius of solar-like oscillators  \citep{2014ApJS..210....1C}.
	
	As for the age, we show in Fig. \ref{fig:HH} that the situation is not as good. Indeed, the scaling relations do not constrain this parameter tightly and the age inference is highly sensitive to the input physics of stellar models \citep[see e.g.][]{2014EAS....65...99L}. With SPInS, we find a mean difference of $28$ per cent on age for the ten stars while the mean difference obtained with the pipelines in \citet{2016A&A...592A..14R} is of $23$ per cent. We get a maximum difference on age of $143$ per cent for Blofeld. Even if, in this particular study, the ages are very well recovered by SPInS for seven artificial stars out of ten, real stars by far host much more subtle physical processes than stellar models are able to describe. Therefore, individual oscillation frequencies  if available, or some combinations thereof, should always be preferred to the seismic indices when a precise and accurate {age} estimate  is {being} sought \citep[see for instance the study of the CoRoT target \object{HD 52265} by][]{2014A&A...569A..21L}.
	
	We also would like to point out that the scaling relations are much less efficient at predicting masses, radii, and ages when the luminosities of the stars are not known. This was checked by applying SPInS to the ten stars using only $T_\mathrm{eff}$, [Fe/H], $\nu_\mathrm{max, obs}$, and $\langle\Delta\nu\rangle_\mathrm{obs}$ as input constraints and removing the constraint on luminosity. In that case, the mean errors on the predicted masses, radii, and ages are higher, with values of $6$, $2$, and $64$ per cent respectively and maximum errors of $19$, $7$ and $300$ per cent for Blofeld. This favours combining all possible classical and asteroseismic parameters to characterise stars, and reinforces the need for precise luminosities from the Gaia mission and radii from interferometry or eclipsing binary light curves.
	
	\subsubsection{Real stars: the Kepler LEGACY sample}
	
	In the same vein, we now consider 66 stars belonging to the Kepler seismic LEGACY sample \citep[e.g.][]{2017ApJ...835..172L}. Each star has at least 12 months of Kepler short-cadence data. Therefore, these stars are among the solar-like oscillators observed by Kepler that have the highest signal-to-noise ratios. As a consequence, their individual oscillation frequencies inferred by \citet{2017ApJ...835..172L} are among the most precise to-date {for solar-like pulsators} while their effective temperatures and metallicity are also available.
	\citet{2017ApJ...835..173S} performed a thorough modelling of the stars, with different optimisation methods implemented in six pipelines. All pipelines took into account the complete set of oscillation frequencies, either individual frequencies or frequency separation ratios, or a combination thereof \citep[see ][for details]{2017ApJ...835..173S}.
	
	We have analysed these stars with SPInS in a simplified way, by considering as observational constraints $T_\mathrm{eff}$, [Fe/H], $\log g$, $\langle\Delta\nu\rangle$, and $\nu_\mathrm{max}$. We used the solar-scaled non-canonical BaSTI grid including convective core overshooting. As for the priors, we adopted the two-slopes IMF from Eq. \ref{eq:priori_imf} with \citet[][]{2013pss5.book..115K}'s coefficients (Eq. \ref{eq:priori_slopeimf2}), a uniform truncated SFR (Eq. \ref{eq:priori_sfr}), and a flat prior on the MDF. We then compared SPInS inferences with those reported in \citet{2017ApJ...835..173S}.
	
\begin{figure*}[ht]
		\begin{center}
			\resizebox{1.0\hsize}{!}{\includegraphics{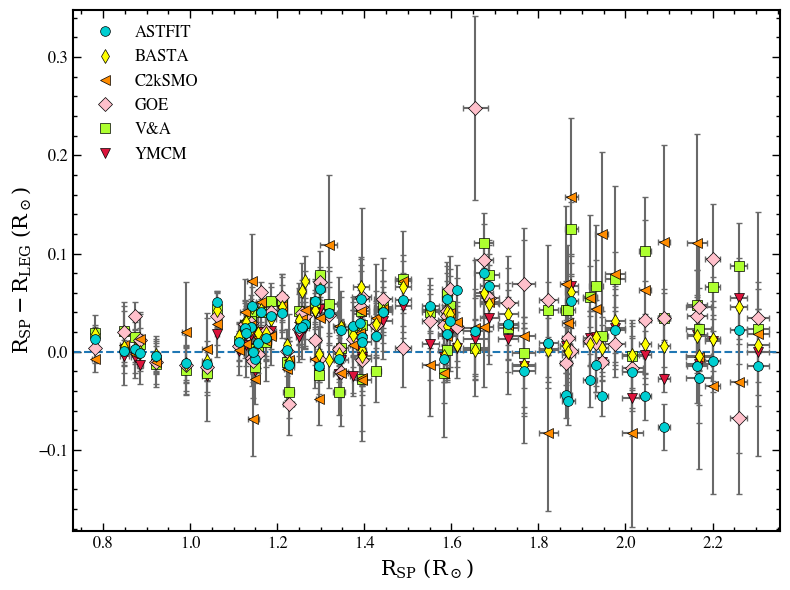}\includegraphics{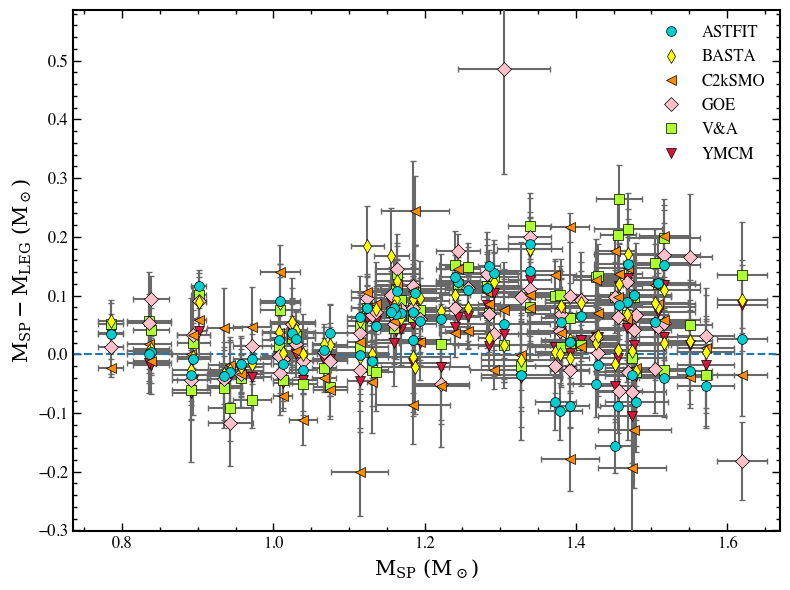}}
			\caption{Left figure: Residuals ($R_\mathrm{SP}-R_\mathrm{LEG}$) between the radius of each of the 66 stars in the Kepler LEGACY sample delivered by SPInS ($R_\mathrm{SP}$) and the values $R_\mathrm{LEG}$ obtained from full optimisations of stellar models based on their individual oscillation frequencies by different pipelines \citep[see e.g.][]{2017ApJ...835..173S}. Right figure: Same comparisons for the mass.  SPInS inference is based on observational constraints on $T_\mathrm{eff}$, [Fe/H], $\log g$, $\langle\Delta\nu\rangle$, and $\nu_\mathrm{max}$. The position of {KIC 7771282}, as modelled by the GOE pipeline is the pink diamond outlier at  $R_\mathrm{SP}\approx\ 1.66 R_\odot$ ($R_\mathrm{LEG, GOE}\approx 1.4 R_\odot$) and $M_\mathrm{SP}\approx\ 1.3 M_\odot$ ($M_\mathrm{LEG, GOE}\approx 0.8 M_\odot$). Symbol colours and shapes identify the pipelines as explained in the legend.}
			\label{fig:LegacyMR}
		\end{center}
	\end{figure*}

	In {the left panel of} Fig.~\ref{fig:LegacyMR}, 
	{we show the residuals ($R_\mathrm{SP}-R_\mathrm{LEG}$) between} the radius of each star inferred by SPInS ($R_\mathrm{SP}$) and the values $R_\mathrm{LEG}$ obtained from full optimisations as reported in \citet[][]{2017ApJ...835..173S}. {In the right panel of Fig.~\ref{fig:LegacyMR}, we show the residuals  ($M_\mathrm{SP}-M_\mathrm{LEG}$)  between the masses.} If we exclude the result of the GOE pipeline for star \object{KIC 7771282} (shown by pink diamonds at {$R_\mathrm{SP}\approx\ 1.66 R_\odot$,} $R_\mathrm{LEG, GOE}\approx 1.4 R_\odot$ and at $M_\mathrm{SP}\approx\ 1.3 M_\odot$, $M_\mathrm{LEG, GOE}\approx 0.8 M_\odot$ in the panels of Fig.~\ref{fig:LegacyMR}), which is well outside the range found by the others, maximum differences on the radius between SPInS and the six pipelines, over 66 stars, range from 3.5 to 9 per cent, while mean differences are in the range of 1.3-2.5 per cent. As for the mass, maximum differences are in the range of 16-26 per cent, while mean differences are in the range of 5-6.5 per cent. Finally, for the ages, we find larger mean differences ranging from 25 to 32 per cent. To get a clearer picture, if we consider the objectives of the PLATO mission \citep{2014ExA....38..249R}, that is to reach uncertainties of less than 2 per cent on the radius, 10 per cent on the mass, and 10 per cent on the age of an exoplanet host-star to be able to characterise its exoplanet correctly, there are three stars for which SPInS's radius is outside the interval corresponding to the extreme values provided by the six pipelines by more than 2.5 per cent, no star with a mass outside the pipeline mass range by more than 10 per cent, and 16 stars with SPInS's age outside the pipeline age range by more than 15 per cent. 
	
	Therefore, overall, SPInS's results compare rather satisfactorily with tight asteroseismic inferences of stellar radii and masses even if  $\langle\Delta\nu\rangle_\mathrm{sc}$ and $\nu_\mathrm{max,\,sc}$ do not perfectly represent observations. Indeed, the ability of  $\langle\Delta\nu\rangle_\mathrm{sc}$ to reproduce $\langle\Delta\nu\rangle_\mathrm{obs}$ relies on asymptotic developments and, as estimated by \citet{2011A&A...530A.142B,2013ASPC..479...61B},  on the main sequence overall departures between the two can reach up to 5 per cent. Regarding the seismic index $\nu_\mathrm{max,\,sc}$, the relation $\nu_\mathrm{max,\,sc}-\nu_\mathrm{max,\,obs}$ is not straightforward. It is intimately related to the acoustic cut-off frequency, a function of  $T_\mathrm{eff}$ and $\log g$, but other properties of the surface layers also play a role which generates biases, in particular on the main sequence because of  $T_\mathrm{eff}$ dispersion (see \citealt{1992MNRAS.255..603B, 2008A&A...485..813C, 2011A&A...530A.142B, 2013ASPC..479...61B} for details). 
	As a consequence, as pointed out by \citet{2015MNRAS.452.2127S}, the use of scaling relations when individual frequencies are unavailable may lead to  wrong estimates of the radius and the mass. Concerning the ages, predictions based on seismic relations are very coarse and SPInS's ages are often far from being a tight inference, which lends more credence to  the words of caution of \citet{2014ApJS..210....1C} regarding age estimates based on scaling relations.
	
	To summarise, SPInS can be a very efficient tool for ensemble asteroseismology \citep[e.g.][]{2013ARA&A..51..353C}, that is to size up, weigh, and age-date large samples of stars with observed values of  $\langle\Delta\nu\rangle_\mathrm{obs}$ and $\nu_\mathrm{max,\,obs}$. However, the reliability of the results will depend on the mass and evolutionary state of the studied stars, and as such must carefully {be} assessed before {deriving} any conclusion. Furthermore, SPInS can also be very useful in providing first estimates of a {star's} mass and age, to be used as initial conditions for refined optimisation methods based on individual oscillation frequencies. 
	
	We point out that the asteroseismic indices used in SPInS do not necessarily have to be obtained via scaling relations.  Instead, they can be calculated with stellar oscillation codes, thus increasing their accuracy, and supplied along with global properties of the model.  This opens up the possibility of using many different types of seismic indicators such as the large and small frequency separations for pressure modes, frequency separation ratios, and the period spacings for gravity modes.  With these quantities, deciphering the age or evolutionary state in advanced stages is accessible with SPInS, thus paving the way to further in-depth studies based on individual oscillation frequencies.
	
	\section{Parameter determination for coeval stars}
	\label{sect:Ensembles}
	
	One attractive feature of the SPInS tool is its ability to deal with stellar groups sharing some common properties. Well-known examples are stars that are members of stellar clusters or of binary systems and for which it can be assumed that they share the same age and initial chemical composition. We illustrate the performances of SPInS with a couple of  study cases below.
	
	\subsection{Stellar clusters}
	
	\begin{figure}[ht]
		\begin{center}
			\resizebox{\hsize}{!}
			{\includegraphics{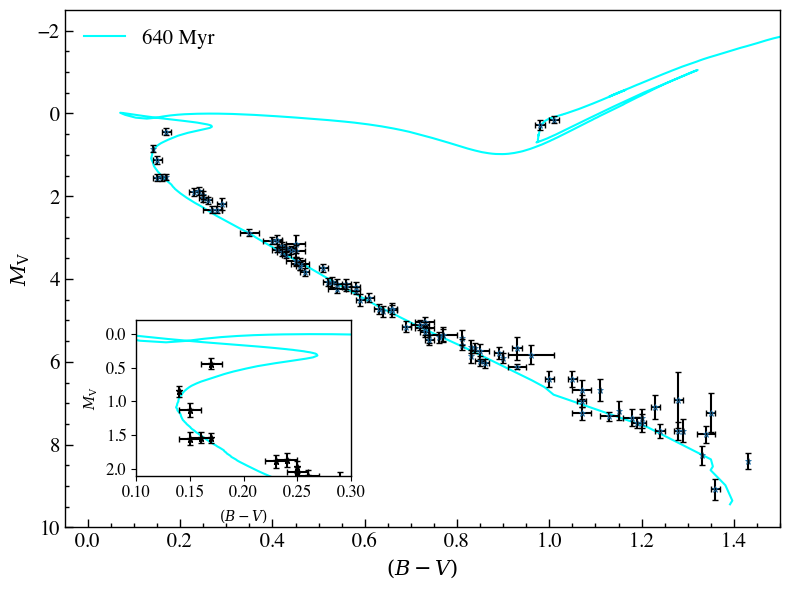}}
			\caption{Colour-magnitude diagram of the 92 stars members of the Hyades. Superimposed 
				is a non-canonical solar-scaled isochrone of $640$ Myr and $\mathrm{[M/H]}=+0.094$ generated by SPInS based on the solution.
				The inset is a zoomed-in portion of the turn-off region.}
			\label{fig:HR_Hyades}
		\end{center}
	\end{figure}
	
	\begin{figure}[ht]
		\begin{center}
			\resizebox{\hsize}{!}
			{\includegraphics{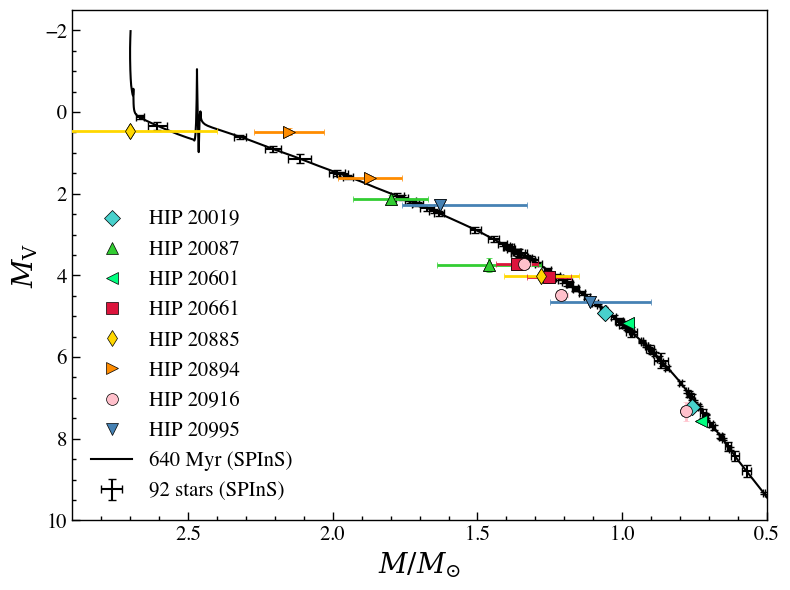}}
			\caption{M--L relation of the Hyades open cluster. Black points with error bars show the position (mass, absolute $V$-magnitude) of 92 Hyades members whose common age ($640 \pm 7$ Myr) and individual masses have been obtained by SPInS. The corresponding SPInS inferred isochrone is plotted in black.
				The positions of 17 stars, members of binary or triple systems  and not included in the 92 star sample, are shown. Their dynamical masses have been inferred directly from observations and are therefore independent of stellar models.}
			\label{fig:ML_Hyades}
		\end{center}
	\end{figure}
	
	As a case study, we consider the Hyades which has long been known to be the nearest open cluster, hosting about 300 members. Hipparcos made it possible to determine secure individual parallaxes of $\sim300$ Hyades members and the cluster remained the only one {for which this was true} until the delivery of Gaia DR1 \citep[e.g. first data release,][]{2017A&A...601A..19G}. The distance to the centre of mass of the Hyades, as determined by \citet{1998A&A...331...81P} from Hipparcos data, is $46.34 \pm 0.27 $ pc, based on 134 stars within 10 pc of the centre. Later on, \citet{1997ESASP.402..733D} and \citet{2001A&A...367..111D} provided a clean sub-sample of 92 Hyades members built from precise Hipparcos parallaxes, and $V$-magnitude and $(B-V)$ colour index from photometric ground-based measurements. \citet{2001A&A...374..540L} then estimated the age of the cluster from eye-fitting of isochrones calculated with a metallicity of $\mathrm{[Fe/H]=0.14\pm 0.05}$ as determined by \citet{1997ESASP.402..687C}. They derived an age of $\simeq 625$ Myr ($\simeq 550$ Myr) on the basis of stellar models with (respectively without) convective core overshooting calculated with the Cesam code \citep{2008Ap&SS.316...61M}.
	
	Starting from the same observed sample as used by \citet{2001A&A...374..540L}, we re-inferred the properties of the cluster {using} SPInS.  As for the priors, we took the  IMF from \citet[][]{1955ApJ...121..161S} as given by   Eqs. \ref{eq:priori_imf1} and \ref{eq:priori_slopeimf1},  a uniform truncated SFR (Eq. \ref{eq:priori_sfr}), and a flat prior on the  MDF. The colour-magnitude diagram of the Hyades is presented in Fig.~\ref{fig:HR_Hyades}. The age of the cluster is found to be $640 \pm 7$ Myr on the basis of BaSTI stellar models including convective core  overshooting, while it is of  $543 \pm 6$ Myr if models with no  overshooting are used instead. These results are in excellent agreement with those obtained by \citet{2001A&A...374..540L}. Furthermore, the common metallicity of the cluster stars inferred by SPInS is $\mathrm{[M/H]}=0.094 \pm 0.003$ with overshooting, while it is $\mathrm{[M/H]}=0.092 \pm 0.003$ without overshooting. These values are lower than the observed value $0.14\pm0.05$, but remain within the error bars.
	
	As a by-product, the age-dating of the 92 coeval cluster members has also provided inferences on their individual masses. Therefore, we drew the mass-luminosity relation (hereafter M--L relation) of the cluster as shown  in Fig.~\ref{fig:ML_Hyades}. Furthermore, it is possible to inter-compare  independently this relation with the observed one. Indeed, there are several  binary systems in the Hyades whose dynamical masses have been derived from orbit analysis. We have inventoried eight binary systems. Five of them (\object{HIP 20019}, \object{HIP 20087}, \object{HIP 20661}, \object{HIP 20885}, \object{HIP 20894}) have been studied for several decades now. Their  M--L relation has been compared with results of stellar models by \citet{2001A&A...374..540L} and revisited by \citet{2019ApJ...883..105T}. Also, a few years ago, \citet{2015A&A...573A.138B} detected solar-like oscillations in the giant star {HIP 20885}A ($\theta^1$ Tau A). From the oscillation power spectrum, they inferred the large frequency separation and frequency at maximum power which allowed them to improve the precision on the star's mass. More recently, the properties of two new systems have been derived by G. Torres and collaborators:  the binary system {80 Tau}, that is \object{HIP 20995} \citep{2019ApJ...883..105T} and the triple system  \object{HIP 20916} \citep{2019ApJ...885....9T}.  Also, \citet{2016MNRAS.455.3303H, 2020MNRAS.496.1355H} obtained the individual masses of the components of \object{HIP 20601}, combining interferometry with the PIONIER instrument  at ESO's VLTI and spectroscopy with the SOPHIE spectrograph at Haute-Provence Observatory. We therefore have in hand 17 stars with known individual masses. Their positions in the M--L plane shown in Fig. \ref{fig:ML_Hyades} fit very well the M--L relation provided by SPInS. Conversely, we ran SPInS with this sample of 17 stars, taking this time their mass, absolute $V$-magnitude, and metallicity as observational inputs with the constraint that they have the same age and metallicity. SPInS provided an age of $615\pm95$ Myr using the solar-scaled non-canonical BaSTI grid. This age, although less precise than the age of $640\pm 7$ Myr derived from the colour-magnitude diagram positions of the 92 Hipparcos stars because based on a smaller sample and a poor coverage of the MS turn-off, is nevertheless in very good agreement with it.
	{SPInS, therefore, offers many possibilities for studying and comparing coeval ensembles and can be very interesting for all kinds of studies of the dynamics and evolution of the Galaxy.}
	
	\subsection{Binary stars}
	\label{sect:binaries}
	
	Binary systems have long provided solid tests of stellar evolution theory, particularly when their components are sufficiently far apart not to undergo mass transfer, since they consist of two stars with different masses that can generally be assumed to share the same age and initial chemical composition.  Different quantities may be accessible {depending on} whether the system is seen as a visual or interferometric binary, a spectroscopic binary (SB), or an eclipsing binary (EB).  Of particular interest are systems that combine the SB with the double-lined character known as SB2 and EB properties, which allows {us to infer} both the individual masses and radii. In this section we apply SPInS to one binary system and compare its inferences with results from the literature.
	
	\subsubsection{AI Phe, a double-lined, eclipsing binary}
	
	\object{AI Phe} is a double-lined, detached eclipsing binary system composed of a main sequence and a subgiant star. A few years ago, \citet{2016A&A...591A.124K} thoroughly characterised the system by obtaining the mass and radius of the components from spectroscopic and photometric measurements. Then, using these masses and radii, effective temperatures and metallicities from the literature together with stellar evolution models, they estimated the age of the system to be $A=4.39 \pm 0.32$ Gyr. Recently, \citet{2020MNRAS.tmp.1795M} revisited the masses and radii on the basis of the light-curves provided by the  TESS mission \citep{2015JATIS...1a4003R}.
	
	Starting from these results, compiled in Table~\ref{tab:AIPhe}, and using the BaSTI solar-scaled non-canonical model grid, a uniform truncated SFR (Eq. \ref{eq:priori_sfr}), and a flat prior both on the  MDF and IMF, SPInS provided a common age for the two stars of $A=4.38 \pm 0.35$ Gyr, which is in excellent agreement with \citet{2016A&A...591A.124K}'s value. However, we point out that \citeauthor{2016A&A...591A.124K}'s analysis is more in-depth since they examined  the effects of  the initial helium abundance {and} the mixing-length parameter for convection {on the results}. Because we used a pre-computed BaSTI grid, we did not have the possibility to make these parameters vary. This may explain why SPInS is able to reproduce the observed masses and radii but not the effective temperatures that are colder than the observed ones {but remain} within the error bars (see Table \ref{tab:AIPhe}). In order to improve the fit, we would have to {run} SPInS  with a stellar model grid {with} more stellar parameters. 
	
	\begin{table}
		\centering
		\caption{SPInS inferences of the properties of the components of {AI Phe}. The observed values are taken from \citet{2016A&A...591A.124K} and \citet{2020MNRAS.tmp.1795M}.}
		\begin{tabular}{lccc}
			\hline \hline
			quantity & observed & inferred\\
			\hline
			$\mathrm{[Fe/H]}$                                            & $   -0.14   \pm   0 .10  $ & $   -0.16   \pm   0.09   $  \\   
			Age (Gyr)                                                    & $   -                    $ & $   4.38    \pm   0.35   $  \\
			\hline
			$M_\mathrm{A} (M_\odot)$                                     & $   1.1938  \pm   0.0008 $ & $   1.1927  \pm   0.0010 $  \\
			$R_\mathrm{A}  (R_\odot) $                                   & $   1.8050  \pm   0.0022 $ & $   1.8055  \pm   0.0020 $  \\
			$T_\mathrm{eff, A}$  (K)                                     & $   6310    \pm   150    $ & $   6266    \pm   119    $  \\
			$ L_\mathrm{A} (L_\odot)$                                   & $   -                    $ & $   4.525   \pm   0.350  $  \\
			$\log g_\mathrm{A}$                                          & $   -                    $ & $   4.001   \pm   0.001  $  \\ 
			$\langle \Delta \nu \rangle_\mathrm{sc, A} (\mu\mathrm{Hz})$ & $   -                    $ & $   60.82   \pm   0.10   $  \\
			$\nu_\mathrm{max, sc, A}                   (\mu\mathrm{Hz})$ & $   -                    $ & $   1086    \pm   11     $  \\
			\hline
			$M_\mathrm{B}  (M_\odot)                                   $ & $   1.2438  \pm   0.0008 $ & $   1.2449  \pm   0.0010 $  \\
			$R_\mathrm{B} (R_\odot)$                                     & $   2.9332  \pm   0.0023 $ & $   2.9329  \pm   0.0020 $  \\
			$T_\mathrm{eff, B}$  (K)                                     & $   5237    \pm   140    $ & $   5114    \pm  54     $  \\
			$ L_\mathrm{B} (L_\odot)$                                    & $   -                    $ & $   5.289   \pm   0.223  $  \\ 
			$\log g_\mathrm{B}$                                          & $   -                    $ & $   3.598   \pm   0.001 $  \\
			$\langle \Delta \nu \rangle_\mathrm{sc, B} (\mu\mathrm{Hz})$ & $   -                    $ & $   30.01   \pm   0.03   $  \\
			$\nu_\mathrm{max, sc, B} (\mu\mathrm{Hz})$                   & $   -                    $ & $   475     \pm   3      $  \\
			\hline  
			\hline
		\end{tabular}
		\label{tab:AIPhe}
	\end{table}	
	
	\section{Conclusion}
	\label{sect:Conclusion} 
	
	We have presented SPInS, a Python-Fortran tool dedicated to the inference of stellar properties  in various
	observational situations. SPInS is a spin-off of AIMS, a sophisticated tool focusing on thorough asteroseismic inferences using stellar models together with their individual oscillation frequencies.
	SPInS is simpler than AIMS in the sense that it only requires standard outputs of stellar models and no {individual oscillation} frequencies to operate. It can be applied to age-date, weigh, and size up stars or groups of stars, as well as to make predictions on their expected global or mean properties such as asteroseismic {indices}, with a considerable gain in computing time compared to AIMS. SPInS {aims to be} user-friendly and can run on any computer cluster or even laptop.
	
	We have first presented the fundamentals of SPInS  as well as its inputs and outputs. As a pre-requisite, SPInS needs to have a stellar evolution model grid available. In the public version of SPInS we provide the solar-scaled and $\alpha$-enhanced canonical and non-canonical BaSTI grids described in Sect.~\ref{sect:models}, but see \citet{2004ApJ...612..168P} and \citet{2006ApJ...642..797P} for extensive descriptions. The grids have been downloaded from the BaSTI website and are available in a format compatible with SPInS. As an option, priors on the parameters to be inferred by SPInS may be provided. {The inputs that need to be provided to SPInS consist in a set of} observational  constraints  {chosen by} the user  {and satisfied by} a star or a group of stars sharing common properties, such as the age. The constraints can be of any kind provided they are available as outputs of the stellar model grid or can be directly derived from them. As output, SPInS provides any unknown stellar property {available in the grid}, including the age,  mass, radius, or seismic indices. Any quantity can be provided either as an input if observed or inferred independently, or as an output if unknown.
	
	In order to present the different outputs of SPInS, such as histograms of the PDF of stellar parameters or the estimators of a given quantity, we have run the tool on a set of fictitious stars spanning a wide area in the H--R diagram. We then validated the SPInS program by comparing its inferences with results {from} the literature. We first showed that SPInS is able to reproduce satisfactorily the ages and masses of more than $10^4$ stars of the GCS11 survey as derived from their absolute magnitudes, effective temperatures, and metallicities by \citet{2011A&A...530A.138C}. We then re-visited the properties of different categories of single stars for which we have access to an extended set of observational constraints, {such} as radii from interferometric measurements or seismic indices. Overall, we obtained results in excellent agreement with what has been published before. Finally, we applied SPInS to the study of coeval stars. As case studies, we took the Hyades open cluster stars and the components of the eclipsing SB2 binary system AI Phe once more showing an excellent agreement with previous results. We therefore release SPInS{\footnote{available at  \url{https://gitlab.obspm.fr/dreese/spins}}} as a public tool in the hopes that it will prove to be useful in deciphering the large quantities of  exquisite data currently available thanks to current (Gaia, CoRoT, Kepler, {TESS}) and future (PLATO) space missions and surveys.
	
	\begin{acknowledgements}
		We thank the referee, Ted von Hippel, for helpful suggestions that improved our manuscript. We thank Misha Haywood and our colleagues of the Scientific Organizing Committee of the 5$^{\mathrm{th}}$ International Young Astronomer School on Scientific Exploitation of Gaia Data to have invited us to pilot a hands-on session on stellar age determination which motivated the development of the SPInS public tool. We thank our colleague Santi Cassisi for allowing us to provide the BaSTI evolutionary tracks on our website, in a format directly readable by SPInS.
	\end{acknowledgements}
	
	\bibliographystyle{aa} % style aa.bst
	\bibliography{article} % your references Yourfile.bib
	
	
\listofobjects
\end{document}